\documentclass[11pt,A4paper]{article}
\usepackage[pdftex,hmargin={2cm,2cm},vmargin={2cm,2cm}]{geometry}
\usepackage[utf8]{inputenc}
\usepackage{amsmath}
\usepackage{amsfonts}
\usepackage{graphicx} 
\usepackage[table]{xcolor}
\usepackage[pdftex,bookmarks=false,colorlinks=true,citecolor=blue]{hyperref}
\usepackage{tabularx}

\newcommand{\R}{\mathbb{R}}

\renewcommand{\d}{{\rm d}}

\newcommand{\EIP}{\textnormal{EIP}}
\renewcommand{\H}{\mathcal{H}}

\renewcommand{\S}{\mathcal{S}}

\newcommand{\II}{\mathbf{I}}
\newcommand{\RR}{\mathcal{R}}
\newcommand{\Spz}{\textnormal{Spz}}
\newcommand{\wild}{\textnormal{wild}}
\newcommand{\Q}{\mathcal{Q}}

\newtheorem{theorem}{Theorem}[section]

\newtheorem{assumption}[theorem]{Assumption}

\title{Epidemiological impacts of age structures on human malaria transmission}
\author{Quentin Richard$^{a,*}$, Marc Choisy$^{b,c}$, Thierry Lefèvre$^d$ and Ramsès Djidjou-Demasse$^{d,e}$}

\date{
    {\small $^{a}$IMAG, Univ. Montpellier, CNRS, France.} \\
	{\small $^{b}$Wellcome Trust Major Overseas Programme, Oxford University Clinical Research Unit, Ho Chi Minh City, Vietnam.} \\
	{\small $^{c}$Centre for Tropical Medicine and Global Health, Nuffield Department of Medicine, University of Oxford, UK.} \\
    {\small $^{d}$MIVEGEC, Univ. Montpellier, IRD, CNRS, Montpellier, France.} \\
   {\small $^{e}$École Polytechnique de Thiès, Sénégal.} \\
	{\small $^*$Author for correspondence: quentin.richard@umontpellier.fr}
 }
\begin{document}

\maketitle

\noindent \textbf{Keywords:} malaria modelling, epidemiology, age-structured models, mosquito survival, vector-borne diseases. \\

\noindent \textbf{Abstract:} Malaria is one of the most common mosquito-borne diseases widespread in tropical and subtropical regions, causing thousands of deaths every year in the world. In a previous paper, we formulated an age-structured model containing three structural variables: (i) the chronological age of human and mosquito populations, (ii) the time since they are infected, and (iii) humans waning immunity (i.e. the progressive loss of protective antibodies after recovery). In the present paper, we expand the analysis of this age-structured model and focus on the derivation of entomological and epidemiological results commonly used in the literature, following the works of Smith and McKenzie. We generalize their results to the age-structured case. In order to quantify the impact of neglecting structuring variables such as chronological age, we assigned values from the literature to our model parameters. While some parameters values are readily accessible from the literature, at least those about the human population, the parameters concerning mosquitoes are less commonly documented and the values of a number of them (\textit{e.g.} mosquito survival in the presence or in absence of infection) can be discussed extensively.

\section{Introduction}

Causing more than 600,000 deaths every year \cite{WHO2022}, malaria is one of the most lethal infectious diseases. Despite the progress towards malaria burden reduction, leading to the decrease in cases and deaths over the last twenty years, achieving elimination in more countries remains a challenge \cite{Tizifa2018}. This is especially true given the slight but non-negligible increase of deaths in 2020 during the Covid-19 pandemic \cite{Weiss2021}. Human malaria is caused by one of 5 plasmodial species: \textit{Plasmodium falciparum}, \textit{P. vivax}, \textit{P. malariae}, \textit{P. ovale} and \textit{P. knowlesi} (with \textit{P. falciparum} being the most pathogenic species infecting humans \cite{KhouryEtAl2018}) that are transmitted by the bites of female \textit{Anopheles} mosquitoes, most commonly from \textit{An. gambiae sensu stricto}, \textit{An. coluzzii}, \textit{An. arabiensis} and species of the \textit{An. funestus} complex in Africa \cite{Sinka2010}. Introduced more than one century ago by Ross \cite{Ross1911}, the first mathematical model for malaria transmission was refined later by MacDonald \cite{MacDonald1957}. Models for vector-transmitted diseases are still a wide subject of study in epidemiology, see \textit{e.g.} \cite{Brauer2019,Inaba2017,LiMartcheva2020} and some references therein. In 2021, the WHO African region accounted for about 95\% of cases and 96\% of deaths globally; while 78.9\% of all deaths in this region were among the youngest population, {\it i.e.} less than 5 years old \cite{WHO2022}. Furthermore, the human infectious reservoir is believed to be mostly children between 5 and 15-year-old \cite{Coalson2018,Felger2012}. Considering different age class for the host population then seems a natural requirement (see \textit{e.g.} \cite{BerettaCapasso2018,ForouzanniaGumel2015}). A refinement of these models are age-structured models, where the chronological age is a continuous variable, since it allows to implement any age-distribution in the host survival instead of simply exponential \cite{VogtGeisse2012,VogtGeisse2013} or also while considering  within-host dynamics \cite{Aguas2008,Tumwiine2008}. In addition, the production of gametocytes within a human host is strongly related to the time post-infection \cite{Djidjou2022}, hence a number of recent studies tracking the time post-infection in their models \textit{e.g.} \cite{CaiMartcheva2017b,DjidjouDucrot2013,Wang2018,WangMartcheva2020}. Finally, the importance of mosquito senescence and the need to include it in models was put forward in \cite{Bellan2010,Styer2007} but only considered recently \cite{Rock2015}. In a previous work \cite{Richard2021}, we studied an age-structured model of the transmission of malaria parasites between mosquitoes and humans, where multiple structuring variables are taken into account: chronological and infection ages of both populations, as well as the time since recovery to describe potential humans waning immunity. In \cite{Richard2021}, well-posedness was first handled, then formulas for the basic reproduction number and vectorial capacity were derived, as well as conditions for backward and forward bifurcations.

In this paper, we extend the analysis of the age-structured model previously developed in \cite{Richard2021}. While the earlier work primarily addressed the well-posedness of the proposed model, the existence of steady states, and the precise derivation of the basic reproduction number and vectorial capacity, our current focus is on deriving entomological and epidemiological results (within the framework of a malaria transmission age-structured model) commonly employed in the literature. This follows the approaches of Smith and McKenzie \cite{SmithEtAl2012,Smith2004,Smith2007}, particularly \cite{Smith2004} where the authors derived equations for various statistics within the ODE framework. Here we generalize these results to the age-structured case and emphasize the importance of both chronological ages and infection ages on malaria transmission and effective malaria control programs. In order to quantify the impact of neglecting structuring variables such as chronological age, we assigned values from the literature to our model parameters. While some parameters values are readily accessible from the literature, at least those about the human population, the parameters concerning mosquitoes are less commonly documented and the values of a number of them (\textit{e.g.} mosquito survival in the presence or in absence of infection) can be discussed extensively.

\section{Description of the model} \label{Sec:model}

\subsection{Model overview}

We remind here of the model introduced in \cite{Richard2021} and the different notations that will be used all along the paper. Let call $S_h(t,a)$ the density of humans of age $a\geq 0$, that are susceptible to the infection at time $t\geq 0$. These individuals can become infected due to bites of infected mosquitoes with the rate $\lambda_m(t,a)$, called the force of infection of mosquitoes to susceptible humans of age $a$. The infected human population is additionally structured by the time since infection, called infection age, with $I_h(t,a,\tau)$ the density at time $t$ of individuals of age $a$ that have been infected for a duration $\tau \geq 0$. A human host of age $a$ and infected for a duration $\tau$ can either recover at the rate $\gamma_h(a,\tau)$, or die from the infection at the rate $\nu_h(a,\tau)$. Upon recovery, $R_h(t,a,\eta)$ is the density at time $t$ of human hosts of age $a$ that recovered at time $ t - \eta\geq 0$. Recovered human hosts lose their immunity at the rate $k_h(\eta)$ and return to the susceptible compartment $S_h$. Death due to natural causes can occur at each step of the infection at the age-dependent rate $\mu_h(a)$. Finally, the human reproduction is assumed to occur at the constant rate $\Lambda_h$ depicting a constant flux of newborns.

\begin{figure}[!h]
\begin{center}\includegraphics[width=.75\linewidth]{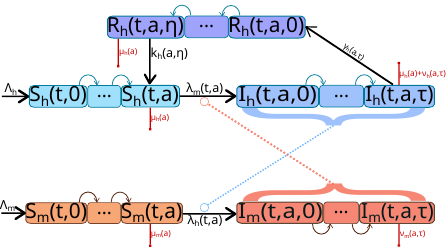}\end{center}
\caption{The model flow diagram. Newborn humans appear with the flux $\Lambda_h$ and are contaminated by infected mosquitoes at rate $\lambda_m$. They recover from the disease at rate $\gamma_h$ and benefit from a temporary immunity that wanes at rate $k_h$. Newborn mosquitoes appear with the flux $\Lambda_m$ and are contaminated by infected humans at rate $\lambda_h$. The death rates are represented by the red lines and are either natural $\mu_k$ or due to infections $\nu_k$ (for $k=h,m$).} \label{Fig:diagram}
\end{figure}

Similarly, we call $S_m(t,a)$ the density of susceptible mosquitoes of age $a$ at time $t$. These mosquitoes become infected upon a blood meal from infected humans at rate $\lambda_h(t,a)$, called the force of infection of humans to mosquitoes with age $a$. Susceptible mosquitoes die with the age-dependent natural death rate $\mu_m(a)$, while infected mosquitoes that have been infected for a duration $\tau$ die at rate $\nu_m(a,\tau)$. Note here that, in order to ensure the positivity of all the parameters, the rate $\nu_m$ is not an additional death rate due to infection as we assumed in \cite{Richard2021}. Indeed, recent experiments show that among old mosquitoes, the survival of infected individuals is higher than that of uninfected counterparts (see \cite{SomeLefevreGuissou2024}) thus leading to $\nu_m(a,\tau) < \mu_m(a)$, at least for old age $a$ and some infection age $\tau$. Finally, as for the human population, the flux of newborn mosquitoes is assumed constant at the rate $\Lambda_m$. The human-mosquitoes infection life cycle is shown in Figure \ref{Fig:diagram} and the infection process is described in the next section. 

\subsection{The infection process}

The total number of human and mosquitoes at time $t$ is described as follows
\[
\begin{split}
&N_h(t)=\int_0^\infty S_h(t,a)\d a+\int_0^\infty \int_0^\infty I_h(t,a,\tau)\d a~\d\tau+\int_0^\infty \int_0^\infty R_h(t,a,\eta)\d a~\d \eta,\\
&N_m(t)=\int_0^\infty S_m(t,a) \d a+\int_0^\infty \int_0^\infty I_m(t,a,\tau)\d a~\d\tau.
\end{split}
\]
The force of infection from mosquitoes to human with age $a$ is then given by:
\begin{equation}\label{Eq:lambda_m}
\lambda_m(t,a)=\frac{1}{N_h(t)}\int_0^\infty \int_0^\infty \theta \beta_m(s,\tau)I_m(t,s,\tau)\d s~\d\tau
\end{equation}
so that $S_h(t,a) \lambda_m(t,a)$ describes the number of newly infected humans with age $a$ at time $t$. It consists of the probability that a human with age $a$ encountered by a mosquito is susceptible $\frac{S_h(t,a)}{N_h(t)}$ and the infection efficiency of the mosquito population $\int_0^\infty \int_0^\infty \theta \beta_m(s,\tau)I_m(t,s,\tau)\d s~\d\tau$. The latter efficiency takes into account: (i) $\theta$ the number of humans bitten by mosquitoes by unit of time and (ii) $\beta_m(s,\tau)$ the probability of parasite transmission from an infected mosquito individual (with age $a$ and which is infected since a time $\tau$) to a human individual. Similarly, the force of infection from a human individual to mosquitoes with age $a$ is given by:
\begin{equation}\label{Eq:lambda_h}
    \lambda_h(t,a)=\frac{1}{N_h(t)} \int_0^\infty \int_0^\infty \theta \beta_h(s,\tau)I_h(t,s,\tau)\d s~\d\tau
\end{equation}
where $\beta_h(s,\tau)$ is the probability of parasite transmission from an infected human with age $s$ and infected since a time $\tau$ to any mosquitoes for each bite.

\subsection{The mathematical model}

Based on the above notations, the model considered in this paper reads as:
\begin{equation}\label{Eq:Model}
\left\{
\begin{array}{rcl}
\left(\frac{\partial}{\partial t}+\frac{\partial}{\partial a}\right)S_h(t,a)&=&\int_0^\infty k_h(\eta) R_h(t,a,\eta)\d \eta-\mu_h(a) S_h(t,a)-S_h(t,a)\lambda_m(t,a) \\
\left(\frac{\partial}{\partial t}+\frac{\partial}{\partial a}+\frac{\partial}{\partial \tau}\right)I_h(t,a,\tau)&=&-\left(\mu_h(a)+\nu_h(a,\tau)+\gamma_h(a,\tau)\right)I_h(t,a,\tau), \vspace{0.1cm} \\
\left(\frac{\partial}{\partial t}+\frac{\partial}{\partial a}+
\frac{\partial}{\partial \eta}\right)R_h(t,a,\eta)&=&-(\mu_h(a)+k_h(\eta))R_h(t,a,\eta), \vspace{0.1cm} \\
\left(\frac{\partial}{\partial t}+\frac{\partial}{\partial a}\right)S_m(t,a)&=&-\mu_m(a)S_m(t,a)-S_m(t,a)\lambda_{h}(t,a), \\
\left(\frac{\partial }{\partial t}+\frac{\partial }{\partial a}+\frac{\partial }{\partial \tau}\right)I_m(t,a,\tau)&=&-\nu_m(a,\tau)I_m(t,a,\tau), \\
\end{array}
\right.
\end{equation}
for each $(t,a,\tau,\eta)\in (0,\infty)^4$ and is associated to the following boundary conditions:
\begin{equation}\label{Eq:Bound_cond}
\left\{
\begin{array}{rclll}
S_h(t,0)&=&\Lambda_h, &S_m(t,0)=\Lambda_m,\\
I_h(t,a,0)&=&S_h(t,a)\lambda_m(t,a), &I_h(t,0,\tau)=0, \vspace{0.1cm} \\
R_h(t,a,0)&=&\int_0^\infty \gamma_h(a,\tau)I_h(t,a,\tau) \d \tau, &R_h(t,0,\eta)=0, \vspace{0.1cm} \\
I_m(t,a,0)&=&S_m(t,a)\lambda_{h}(t,a), &I_m(t,0,\tau)=0
\end{array}
\right.
\end{equation}
and initial conditions (at $t=0$):
\begin{equation}\label{Eq:Ini_cond}
\left\{
\begin{array}{rclrlrl}
S_h(0,a)&=&S_{h,0}(a), &&I_h(0,a,\tau)&=&I_{h,0}(a,\tau), \qquad R_h(0,a,\eta)=R_{h,0}(a,\eta), \\
S_m(0,a)&=&S_{m,0}(a), &&I_m(0,a,\tau)&=& I_{m,0}(a,\tau), 
\end{array}
\right.
\end{equation}
for each $(a,\eta,\tau)\in \R_+^3$. The notations of all variables and parameters are summarized in Table \ref{Tab-parameters}, as well as the biological meaning and the references used or discussed for the parameterization. Such age-structured model recover the classical model with SIRS compartments for humans and SI compartments for mosquitoes. As we will see in Section \ref{Sec:Comparison}, assuming piecewise functions can reveal the exposed compartment, yielding a SEIRS model if among humans or SEI model if among mosquitoes. Furthermore, Model \eqref{Eq:Model}-\eqref{Eq:Bound_cond} allows for the explicit consideration of the asymptomatic stage, typically expressed during the infection process. This is of particular significance since asymptomatic malaria infections are highly prevalent in endemic areas and only a small percentage of infections will exhibit clinical symptoms (mostly young individuals). We show in Section \ref{Sec:SAIRmodel} how Model \eqref{Eq:Model}-\eqref{Eq:Bound_cond} can explicitly highlight the asymptotic stage before infection as for some classical model formulation of SAIR type. We mention here that since antigenic variation is a major driver of malaria dynamics, some papers \cite{Aguas2008} or \cite[Section 8.3]{Inaba2017} considered cases where the epidemiological parameters differ between reinfected individuals and individuals infected for the first time.

\begin{table}[!htp]
	\begin{small}
		\begin{tabular}{clll}
			\hline \hline
			Category & Description & Unit & References \\
			\hline \hline
			Notations &    & &   \\
			\hline
				$t$ & Time & Tu &  \\
					$a$ & Chronological age & Tu & \\
			$\tau$ & Time since infection  & Tu & \\
					$\eta$ & Time since recovery for human  & Tu & \\
			\hline
			State variables &  &  &  \\
			\hline
			$S_h(t,a), S_m(t,a)$ & Susceptible humans and mosquitoes & No unit & \\ 
			$I_h(t,a,\tau), I_m(t,a,\tau)$ & Infected humans and mosquitoes & No unit & \\
			$N_h(t), N_m(t)$ & Total human and mosquito populations & No unit &\\
			\hline 
			Initial conditions &    &  & \\
			\hline
			$S_{h,0}(a)$ & Initial human susceptible population & No unit & \cite{INSD-RGPH2006-T2,INSD2015} \\ 
			$S_{m,0}(a)$ & Initial mosquitoes population & No unit & Varying \\
			\hline 
			Parameters & & \\
			\hline
			$\Lambda_h$ & Human recruitment rate & Tu$^{-1}$ & \cite{INSD-RGPH2006-T2,INSD2015} \\
		    $\Lambda_m$ & Mosquitoes recruitment rate & Tu$^{-1}$ & Varying \\
        	$\mu_h(a)$ & Human death rate & Tu$^{-1}$ & \cite{INSD-RGPH2006-T7} \\
		    $\mu_m(a)$ & Mosquitoes death rate & Tu$^{-1}$ & \cite{SomeLefevreGuissou2024,Rock2015,Styer2007} \\
        	$\nu_h(a,\tau)$ & Human death rate induced by the infection & Tu$^{-1}$ & \cite{ChitnisCushing2008,ASS2018,Molineaux80} \\
			$\nu_m(a,\tau)$ & Infected mosquitoes death rate & Tu$^{-1}$ & \cite{SomeLefevreGuissou2024} \\
			$\gamma_h(a,\tau)$ & Recovery rate of human infections & Tu$^{-1}$ & \cite{Bekessy76,Molineaux80} \\
			$k_h(\eta)$& Rate of loss of immunity & Tu$^{-1}$ & \cite{ChitnisCushing2008,Keegan2013} \\
			$\beta_h(a,\tau)$ & Parasite transmission probability from human to mosquitoes & No unit & \cite{Barry2021,Bradley2018,Churcher2013,Eichner2001,SomeLefevreGuissou2024} \\
			$\beta_m(a,\tau)$ & Parasite transmission probability from mosquitoes to human & No unit & \cite{Churcher2017,SomeLefevreGuissou2024} \\
			$\theta$ & Human feeding rate & Tu$^{-1}$ & \cite{ChitnisCushing2006,Molineaux79} \\
				\hline \hline
		\end{tabular}\\
		Tu=time unit; h=humans; m=mosquitoes
	\end{small}
	\caption{Main notations, state variables and parameters of the model.} 
	\label{Tab-parameters}
\end{table}

\section{Derivation of the vectorial capacity and  the basic reproduction number}\label{Sec:Param}

From a public health point of view, the time course of the disease is strongly related to the basic reproduction number, denoted here by $\RR_0$. This allows quantifying the expected number of secondary human (respectively mosquito) infections resulting from a single primary human (resp. mosquito) infection into an otherwise susceptible population. In \cite{Richard2021}, it was shown that the $\RR _0$ for the model \eqref{Eq:Model}-\eqref{Eq:Bound_cond} takes the form
\begin{equation}\label{Eq:R0}
\RR _0= \sqrt{  \RR _0^{h\to m} \times \RR _0^{m\to h}}
\end{equation}
where $\RR _0^{h\to m}$ quantifies the per bite transmission capability from humans to mosquitoes and $\RR _0^{m\to h}$ quantifies the transmission capability from mosquitoes to humans. Note that $\RR _0^{m\to h}$ is also called the vectorial capacity \cite{Garrett69}. More precisely, we have \cite{Richard2021}:
%\begin{align*}
%\RR _0^{h\to m}=&   \int_0^\infty \underbrace{\dfrac{e^{-\int_0^a \mu_h(s)\d s}}{\int_0^\infty e^{-\int_0^a \mu_h(s)\d s}\d a }}_{\text{Prop. human of age $a$}} \int_a^\infty \underbrace{\beta_h(s,s-a) \times  \underset{\substack{\text{\tiny Infectiousness prob. of human} \\ \text{\tiny of age $a$ to age $s$}}} {e^{-\int_a^s (\mu_h(\xi)+\nu_h(\xi,\xi-a)+\gamma_h(\xi,\xi-a))\d \xi}}}_{\substack{ \text{Transmission prob. of infected human}\\ \text{of age $a$ to age $s$}}}  \d s~\d a,
%\end{align*}
\begin{align*}
\RR _0^{h\to m}=&   \int_0^\infty \underbrace{\dfrac{e^{-\int_0^a \mu_h(s)\d s}}{\int_0^\infty e^{-\int_0^{\xi}\mu_h(s)\d s}\d \xi }}_{\text{Prop. human of age $a$}} \int_0^\infty \underbrace{\beta_h(a+\tau,\tau) \times  \overbrace{ {e^{-\int_0^{\tau} (\mu_h(\xi+a)+\nu_h(\xi+a,\xi)+\gamma_h(\xi+a,\xi))\d \xi}}}^{\substack{\text{\tiny Prob. to remain infected}}}}_{\substack{ \text{Transmission prob. of infected human}}}  \d \tau~\d a
\end{align*}
and
\begin{align}\label{Eq:vec_capacity}
\RR _0^{m\to h}=  \int_0^\infty \underbrace{\dfrac{e^{-\int_0^a \mu_m(s){\rm d}s}}{\int_0^\infty e^{-\int_0^a \mu_m(s){\rm d}s} {\rm d}a}}_{\text{Prop. mosquito of age $a$}} \times \RR _0^{m\to h}(a) \d a
\end{align}
where $\RR _0^{m\to h}(a)$ denotes the vectorial capacity of mosquitoes population infected at age $a$ and is explicitly given by 
\begin{align*}
\RR _0^{m\to h}(a)=   \underbrace{\dfrac{\Lambda_m \int_0^\infty e^{-\int_0^a \mu_m(s){\rm d} s} {\rm d}a }{ \Lambda_h \int_0^\infty e^{-\int_0^a \mu_h(s){\rm d}s}{\rm d}a}}_{\text{Mosquito/human ratio}} \times  \theta^2 \times  \int_0^\infty \underbrace{ \beta_m(a+\tau,\tau) \times \overbrace{e^{-\int_0^\tau \nu_m(\xi+a,\xi)d \xi}}^{\substack{\text{\tiny Surv. prob. of infected mosquito}}}}_{\substack{ \text{Transmission prob. of infected mosquito}}} {\rm d}\tau.
\end{align*}
In the following, we will focus on the description of each factor intervening in the decomposition of the transmission capabilities and how it can be determined in practice by using existing data. 

\subsection{Human transmission capability}

\subsubsection{Demographic structure of the human population}\label{Sec:HumanDeath}

The demography of the human population plays an important role in the transmission dynamics of malaria. As said in the introduction, the human infectious reservoir is known to be age-dependant. Here we consider the population of Bobo Dioulasso, the second biggest city in  Burkina Faso where the spread of malaria is important.

\begin{figure}[!h]
\centering
\begin{tabular}{ccc}
(a) & (b) & (c)\\
\includegraphics[width=.33\linewidth]{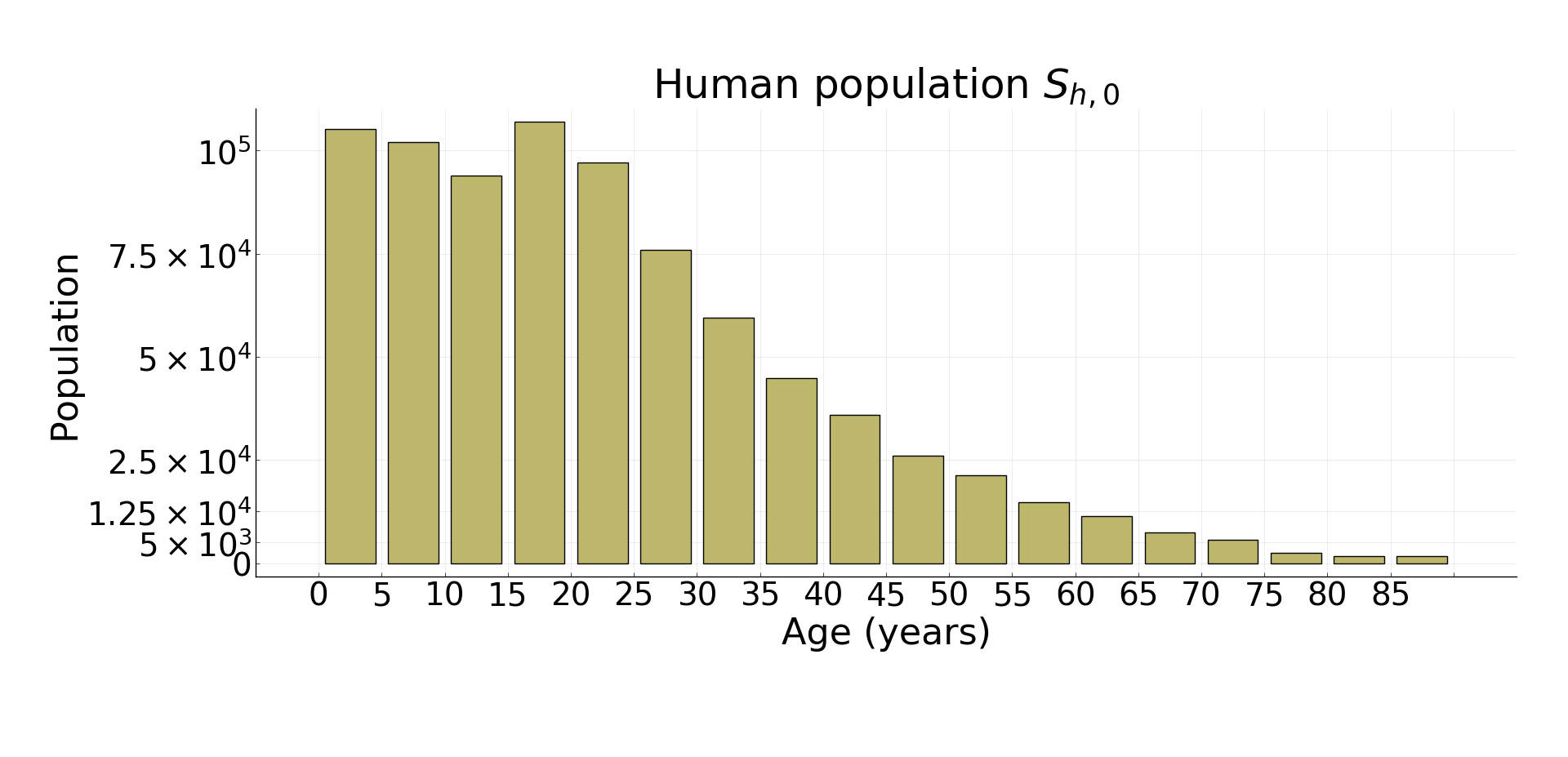} & \hspace{-1cm} \includegraphics[width=.33\linewidth]{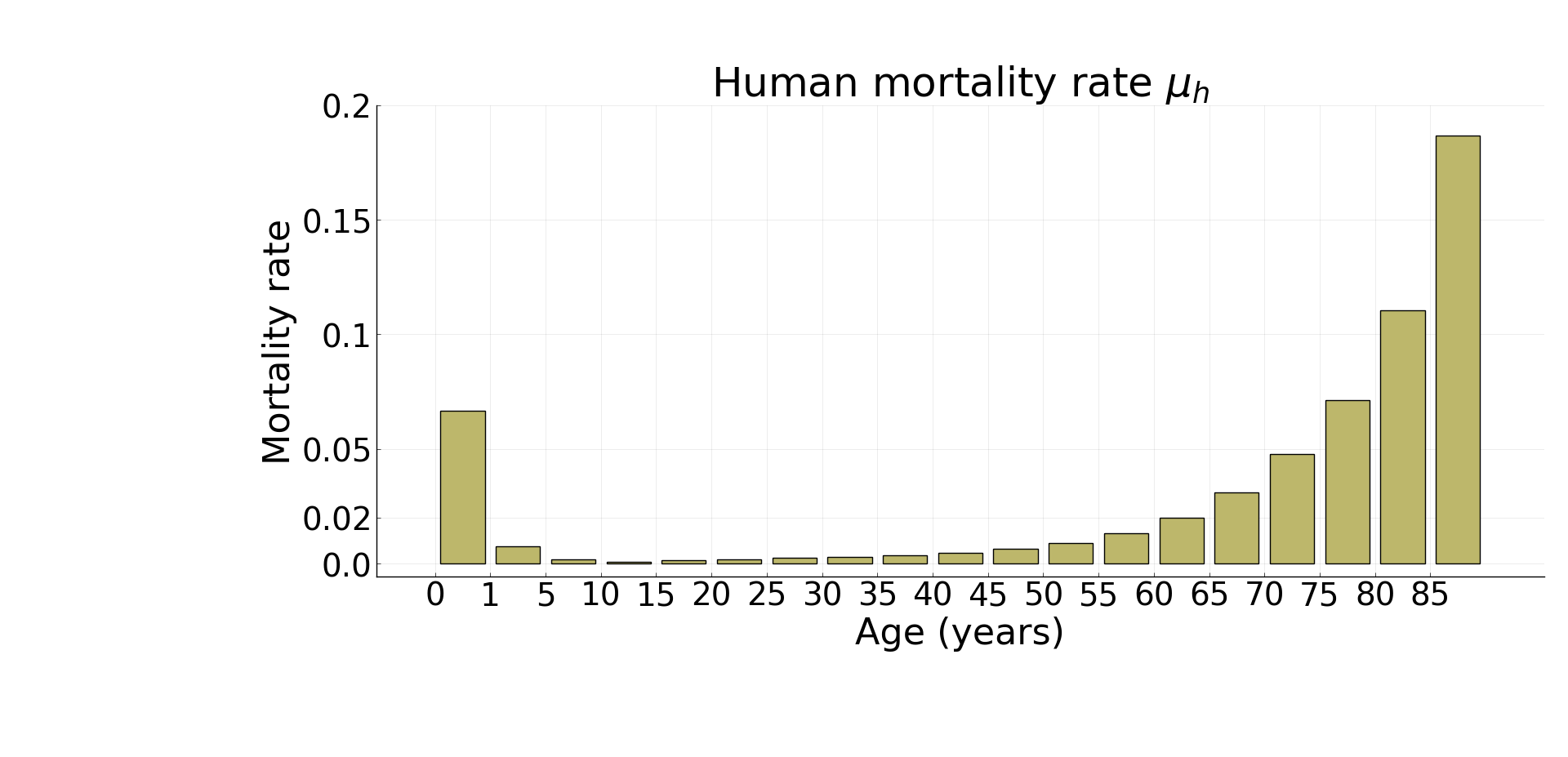} & \includegraphics[width=.33\linewidth]{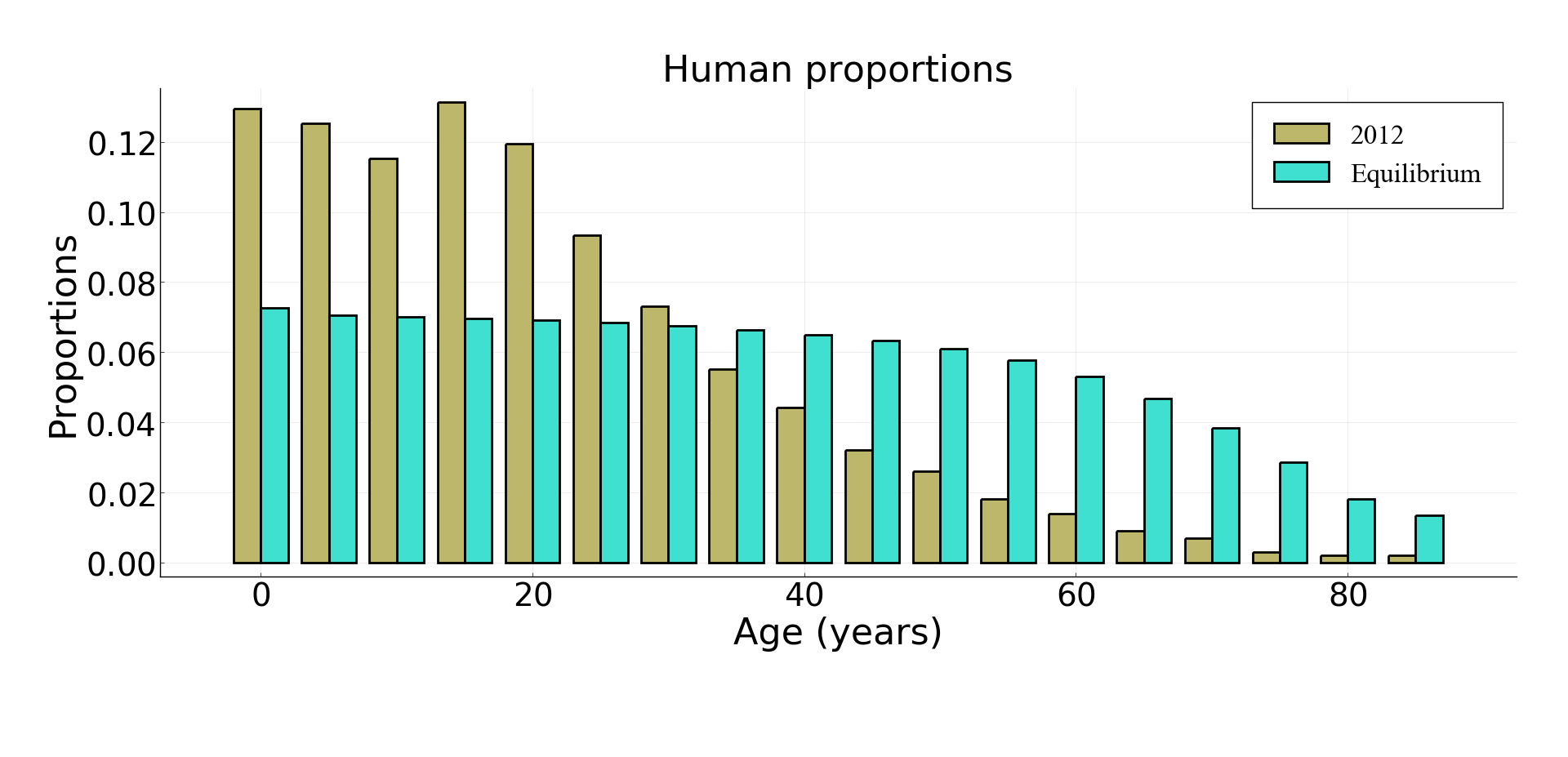}
\end{tabular}
\caption{(a) The population age-structure of Bobo Dioulasso in 2012. (b) The age-dependent mortality rate in Burkina Faso. (c) The expected age-structure of the population, in proportion, in absence of mortality due to malaria, compared to the proportion in 2012.} \label{Fig:pop_Bobo}
\end{figure}

The initial population $S_{h,0}$ corresponds to the population in Bobo Dioulasso in 2012 \cite[Table A.4.6]{INSD2015} with the age-structure of the city \cite[Table 4.6]{INSD-RGPH2006-T2} (see Figure \ref{Fig:pop_Bobo} (a)). The human recruitment rate is chosen as $\Lambda_h=30754$/year corresponding to a birth rate estimated to $3.78\%$ \cite[p.16]{INSD-RGPH2006-T2} within Bobo, with a total population of 813 610 in 2012. The natural human death rate $\mu_h$ is estimated in \cite[Table A.1]{INSD-RGPH2006-T7} (see Figure \ref{Fig:pop_Bobo} (b)).

Considering that malaria mortality $\nu_h$ is negligible compared to the natural mortality, we can estimate the demographic age-structure reached by the population after some time, under the assumption that the mortality does not vary with time. This is given by the function $a\longmapsto\Lambda_h e^{-\int_0^a \mu_h(s)ds}$ which mathematically corresponds to the disease-free equilibrium, where the function $\pi_h^s(a)=e^{-\int_0^a \mu_h(s)ds}$ is the probability for humans to survive from birth to age $a$, in absence of malaria infections. We can then define the proportion of human of age $a$ reached by the population as:
$$P_h(a)=\dfrac{e^{-\int_0^a \mu_h(s)\d s}}{\int_0^\infty e^{-\int_0^\xi \mu_h(s)\d s}\d \xi}=\dfrac{\pi_h^s(a)}{\int_0^\infty \pi_h^s(\xi)d\xi}$$
which appears in the computation of the transmission capability and is represented in Figure \ref{Fig:pop_Bobo} (c).
We can observe a few changes in the demography leading to a larger and older population, with a bigger life expectancy (numerically around 64 years old for about 2 million inhabitants). This can easily be explained by the decreasing of both the mortality and the birth rates over the years.

In practice, the mortality due to malaria infections can indeed be neglected since malaria cause around $4000$ deaths each year in Burkina Faso \cite[Table 4.74]{ASS2018} while the crude mortality rate was estimated to $1.18\%$ in 2006 \cite{INSD2015} corresponding to more than 100 000 deaths each year in Burkina Faso with a total population of 14 millions inhabitants in 2006 \cite{INSD-RGPH2006-T2}. 

\subsubsection{Probability to remain infected}

The time lapse between the infection and the recovery or the death of individuals is a key parameter in the transmission process. Indeed, only infectious individuals can contribute to the spread of malaria parasites. Considering humans getting infected at age $a$, they will remain infected for $\tau$ days with the probability given by
$$\pi^i_h(a,\tau)= e^{-\int_0^\tau (\mu_h(\xi+a)+\nu_h(\xi+a,\xi)+\gamma_h(\xi+a,\xi))\d \xi}$$
revealing the human death rate induced by the infection $\nu_h$ and the recovery rate $\gamma_h$. The mortality rate $\nu_h$ is based on the malaria lethality at age $a$, denoted by $\overline{\nu}_h(a)$. The latter is given in \cite{ASS2018} by computing the number of deaths over the total number of mild \cite[Table 4.73]{ASS2018} and severe \cite[Table 4.74]{ASS2018} malaria cases in the Hauts Bassins region (see Table \ref{Table:Lethality}). 

\begin{figure}[!h]
\centering
\begin{tabular}{cc}
(a) & (b) \\
\includegraphics[width=0.45\linewidth]{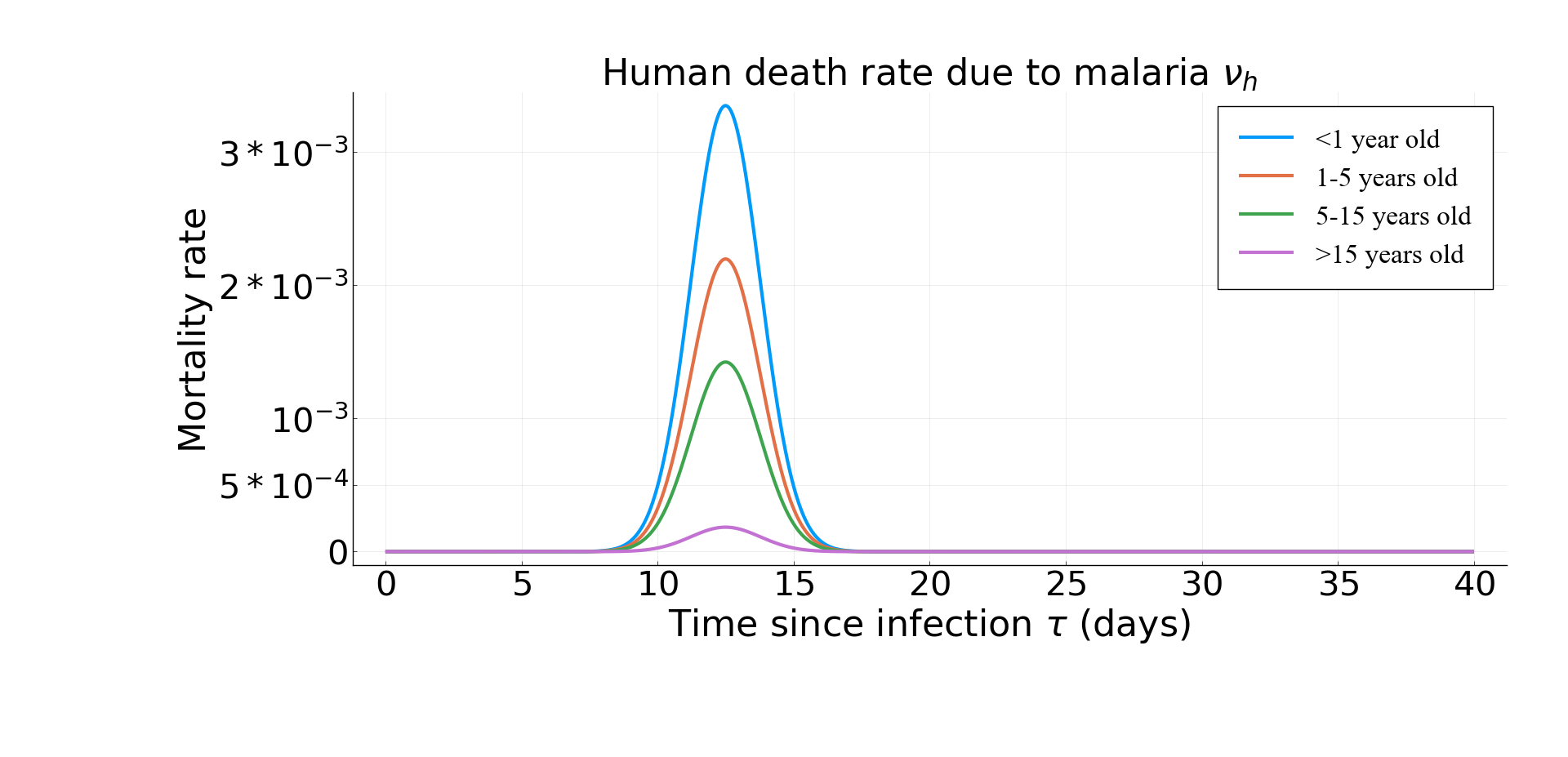} & \includegraphics[width=.45\linewidth]{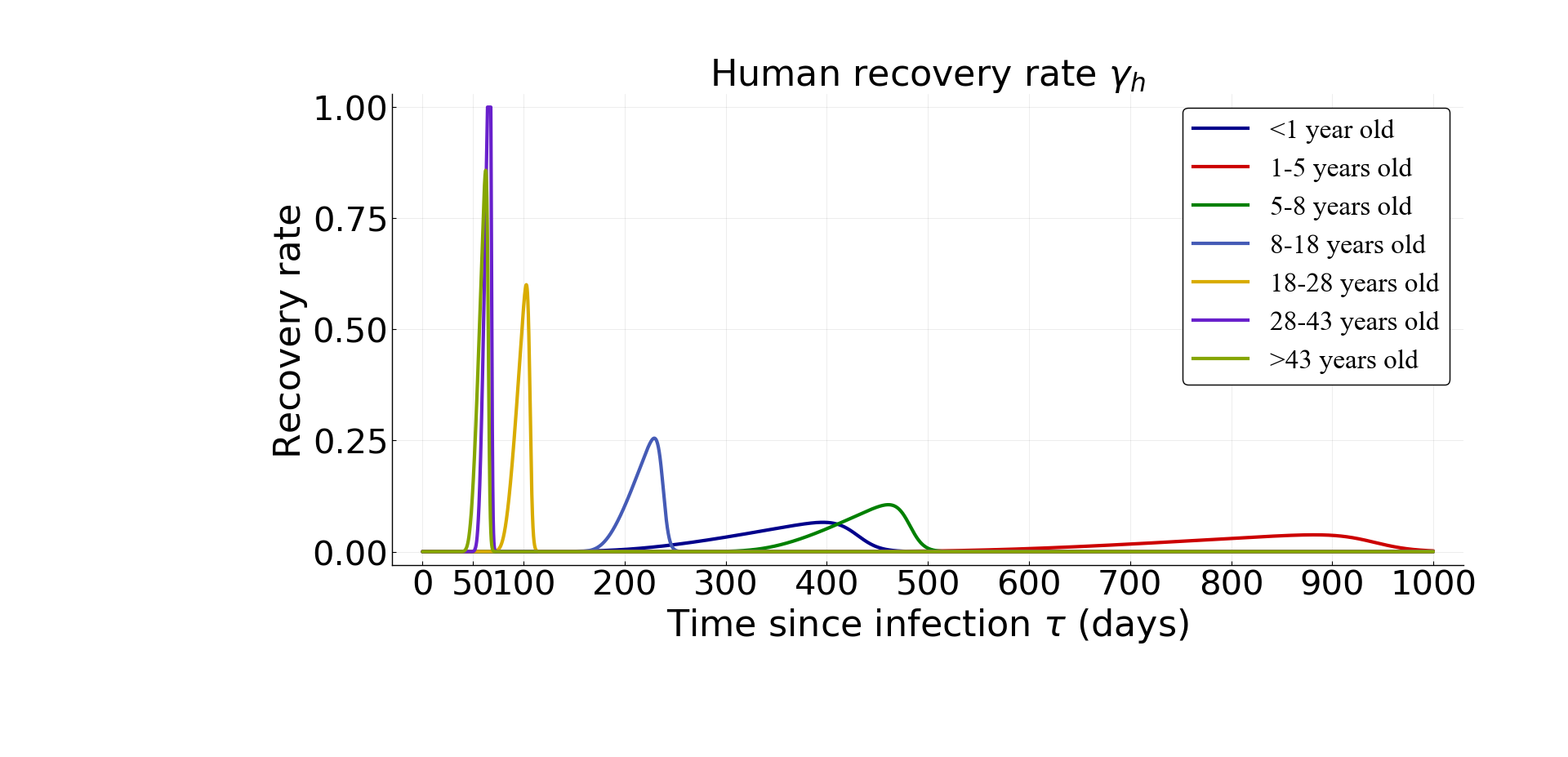}
\end{tabular}
\caption{(a) The mortality rate $\nu_h$ computed such that distribution of deaths due to malaria, according to the time since infection, follows the Gaussian law $\Phi_{\nu}$. (b) The recovery rate $\gamma_h$ depends both on the age and on the time since infection of the infected human. It is computed so that the distribution follows a Gaussian distribution such that 95 $\%$ of the recovery occurs within the ranges showed in Table \ref{Table:Recovery}.} \label{Fig:letha_recov}
\end{figure}

\begin{table}[!htp]
\centering
\begin{tabular}{|c|c|c|c|c|}
    \hline
    Age (years old) & [0,1] & [1,5] & [5,15] & $[15,+\infty)$  \\
    \hline
    Malaria lethality rate & $1.07\times 10^{-3}$ & $7.02\times 10^{-4}$ & $4.55\times 10^{-4}$ & $5.73\times 10^{-5}$ \\
    \hline
\end{tabular}
	\caption{Age-dependent malaria lethality rate for humans} 
\label{Table:Lethality}
\end{table}
The human incubation period is estimated to be approximately 10 days \cite{ChitnisCushing2008, Molineaux80} while death generally occurs within 1 to 5 days after the incubation period, with a mean duration of symptoms until death of $2.8$ days \cite{Greenwood87}. In order to have $95\%$ of the deaths occurring between 10 and 15 days post-infection, we make the assumption that the distribution of the occurrence of the deaths due to malaria, denoted by $\Phi_{\nu}(\tau)$, follows the Gaussian law $\mathcal{N}(12.5,1.276)$ and is normalized to have a total mass equal to one: $\int_0^\infty \Phi_{\nu}(\tau)d\tau=1$, {\it i.e.}
$$\Phi_{\nu}(\tau)=\frac{\exp\left(-\frac{(\tau-12.5)^2}{2\times 1.276^2}\right)}{\int_0^\infty \exp\left(-\frac{(s-12.5)^2}{2\times 1.276^2}\right) \d s}.$$
Finally, the mortality rate $\nu_h(a,\tau)$ of human of age $a$, with $\tau$ days post-infection is chosen as
$$\nu_h(a,\tau)=\overline{\nu}_h(a)\frac{\Phi_{\nu}(\tau)}{1-\overline{\nu}_h(a)F_{\Phi_\nu}(\tau)}$$
 (see Figure \ref{Fig:letha_recov} (a)) where $F_{\Phi_{\nu}}$ denotes the cumulative distribution function of $\Phi_{\nu}$. Under this assumption, the distribution of the deaths due to malaria according to the time since infection numerically satisfies the Gaussian law $\Phi_{\nu}$  when following a cohort of newly infected humans.

\begin{figure}[!h]
\centering
\begin{tabular}{c}
\includegraphics[width=.7\linewidth]{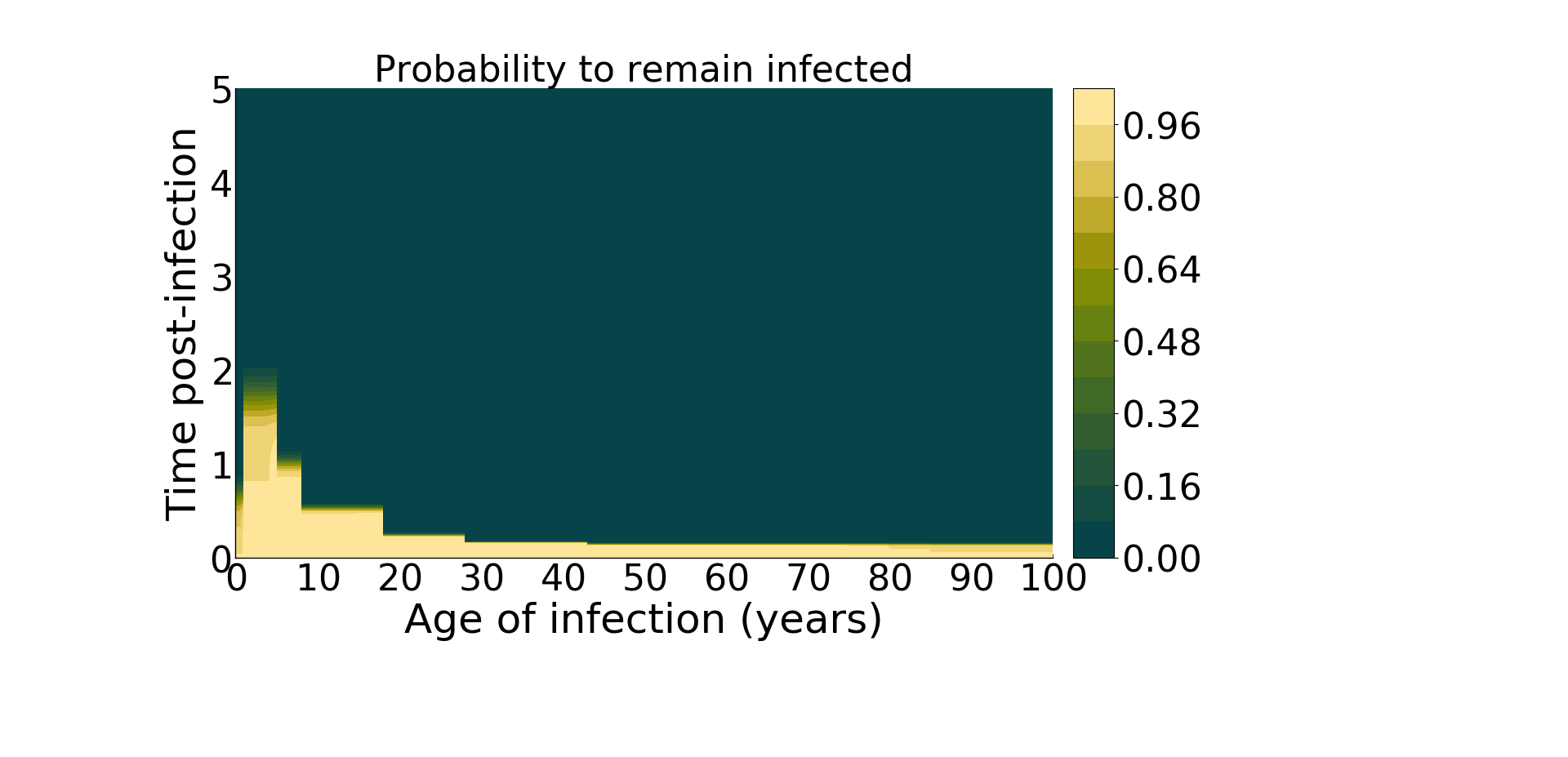}
\end{tabular}
\caption{The probability to remain infected depending on the time post-infection and the chronological age at the time of the infection (both in years).} \label{Fig:remain_infected}
\end{figure}

The time required to clear the parasite was estimated in \cite{Bekessy76,Molineaux80} and is age-dependant (see Table \ref{Table:Recovery}). In order to compute the recovery rate $\gamma_h$, we first make the assumption that the clearance for a human of age $a$ follows the Gaussian distribution $\Phi_{\gamma}(a,\tau)$, according to the time since infection $\tau$, centered in the middle of each range, while we choose the variance so that $95\%$ occurs within this range. We then consider the following recovery rate (see Figure \ref{Fig:letha_recov} (b)):
$$\gamma_h(a,\tau)=\frac{\Phi_{\gamma}(a-\tau,\tau)}{1-F_{\Phi_\gamma}(a-\tau,\tau)},$$
where $F_{\Phi_{\gamma}}$ denotes the cumulative distribution function of $\Phi_{\gamma}$. Note that $a-\tau$ corresponds to the age at which infected humans were contaminated. Finally, the probability $\pi^i_h(a,\tau)$ for an infected human of age $a$ to remain infected for $\tau$ days is illustrated in Figure \ref{Fig:remain_infected}.

\begin{table}[!htp]
\centering
\begin{tabular}{|c|c|c|c|c|c|c|c|}
    \hline
    Age (years) & [0,1] & (1,5] & (5,8] & (8,18] & [18,28]  & (28,43]&  $(43,+\infty)$  \\
    \hline
  Recovery time (days) & [163,345] & [555,714] & [344,400] & [181,204] & [82,92] & [56,61] & [48,55] \\
   \hline
\end{tabular}
	\caption{Age-dependent time for infected humans to recover} 
\label{Table:Recovery}
\end{table}

\subsubsection{Human transmission probability}

\begin{figure}[!h]
\centering
\begin{tabular}{cc}
(a) & (b)\\
\includegraphics[width=.45\linewidth]{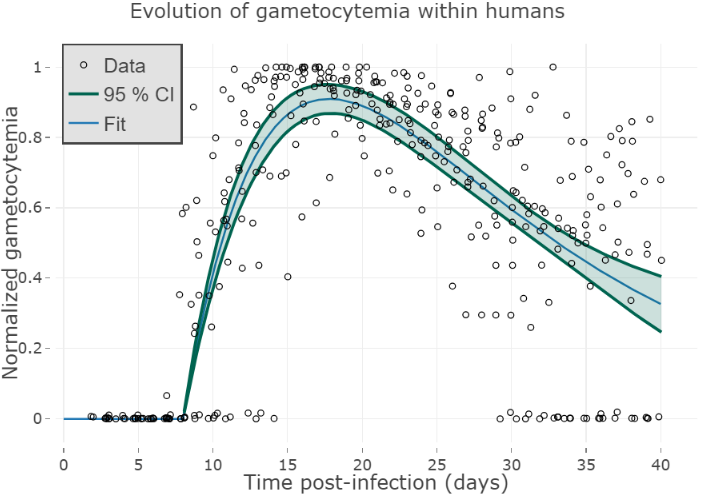} & \includegraphics[width=.5\linewidth]{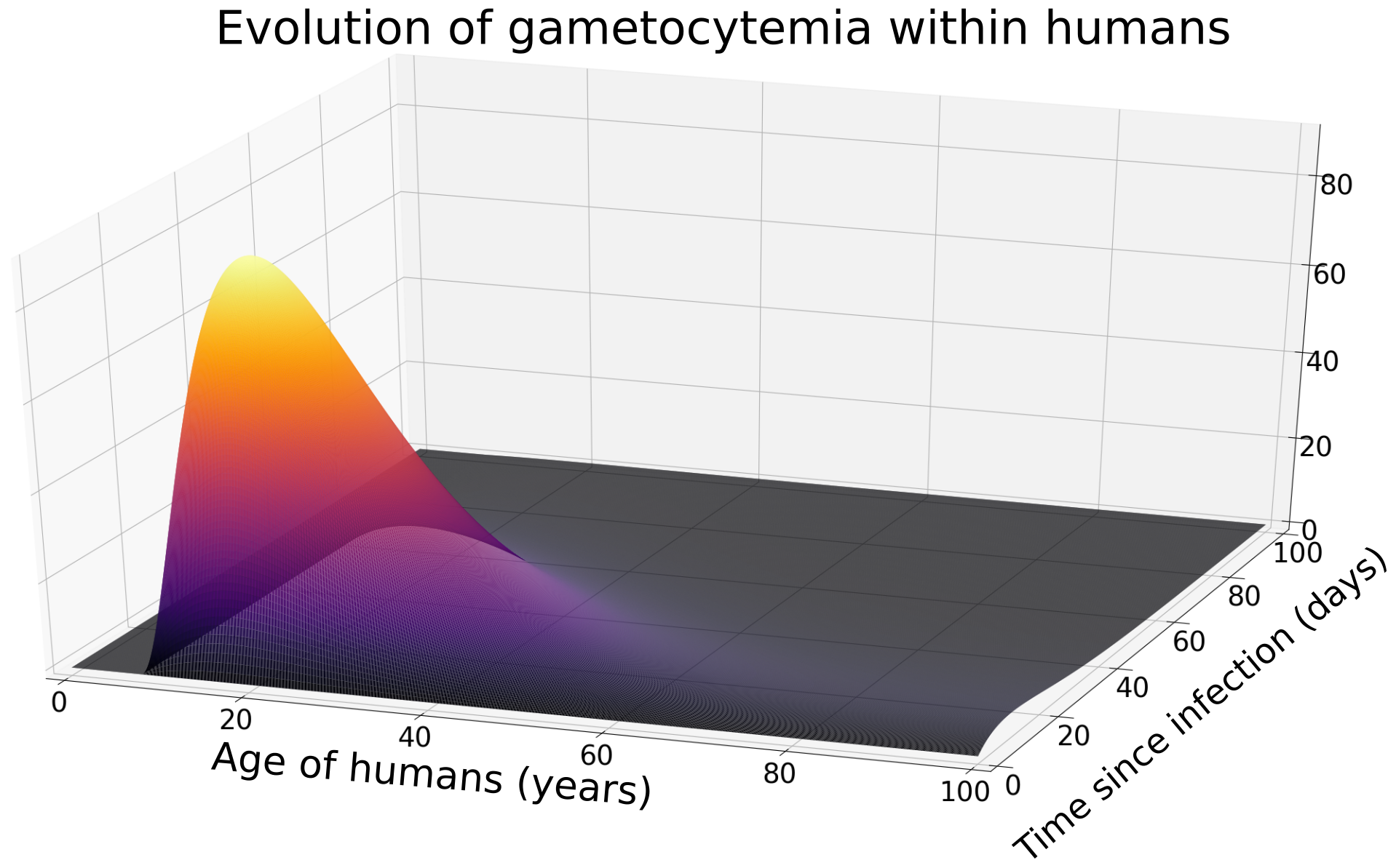}
\end{tabular}
\caption{(a) The evolution of normalized gametocytemia within human population. (b) The evolution of gametocytemia per $\mu$L of blood within human population according to the time post-infection and humans age.} \label{Fig:Gameto_Human}
\end{figure}

Another crucial parameter in the human transmission is the probability for a mosquito to get infected while taking a blood meal on a infected human. For each mosquito bite, we suppose that the probability that a human of age $a$ and infected for $\tau$ days, transmit the infection to the biting mosquito is $\beta_h(a,\tau)$. Such a probability is known to be strongly linked to the gametocytes density \cite{Ahmad2021,Barry2021,Bradley2018,Churcher2013,DaChurcher2015,Eichner2001}. We denote by $G(a,\tau)$ the gametocytes density of an infected human of age $a$, after $\tau$ days of infection. As in \cite{Bradley2018} (see also \cite{Ahmad2021,Barry2021}), we assume that the relationship between gametocyte density $G(a,\tau)$ and the probability of transmission from infected human $\beta_h(a,\tau)$ is given by:
$$
\beta_h(a,\tau)= \alpha_1 \left( G(a,\tau) \right)^{\alpha_2}
$$
where $\alpha_1=0.071 [0.023, 0.175]$ and $\alpha_2=0.302 [0.16, 0.475]$. Moreover, we assume that the gametocytes density follows the function
\begin{equation*}
G(a,\tau)= \left\{\begin{array}{ll}
G_0(a)\times f(\tau-8) & \mbox{if $\tau > 8$,}\\
0 & \mbox{otherwise},
\end{array}\right.
\end{equation*}
where the incubation period within an infected human is approximately 8 days (Figure \ref{Fig:Gameto_Human}(a)), and $G_0(a)$ is the mean number of gametocytes for a human of age $a$ and the function $f$ defined by
$$f:x\longmapsto \xi+(\psi x-\xi) \exp(-\omega x)$$
with some parameters $\xi, \psi$ and $\omega$, gives the evolution of the gametocyte density. It was used in \cite{Churcher2013} to estimate the function $G_0$ as
\begin{equation*}
   G_0(a)=\psi a \exp(-\omega a),
\end{equation*}
where $\xi=0$ and $\psi,\omega$ were respectively estimated to $22.7 [17,32]$ and $0.0934 [0.08,0.11]$.

Finally, the evolution of gametocyte density is fitted on data from 12 patients \cite{Eichner2001}, that were normalized such that the maximum of each gametocyte density is one, due to the variance between individuals.

\begin{figure}[!h]
\centering
\includegraphics[width=1\linewidth]{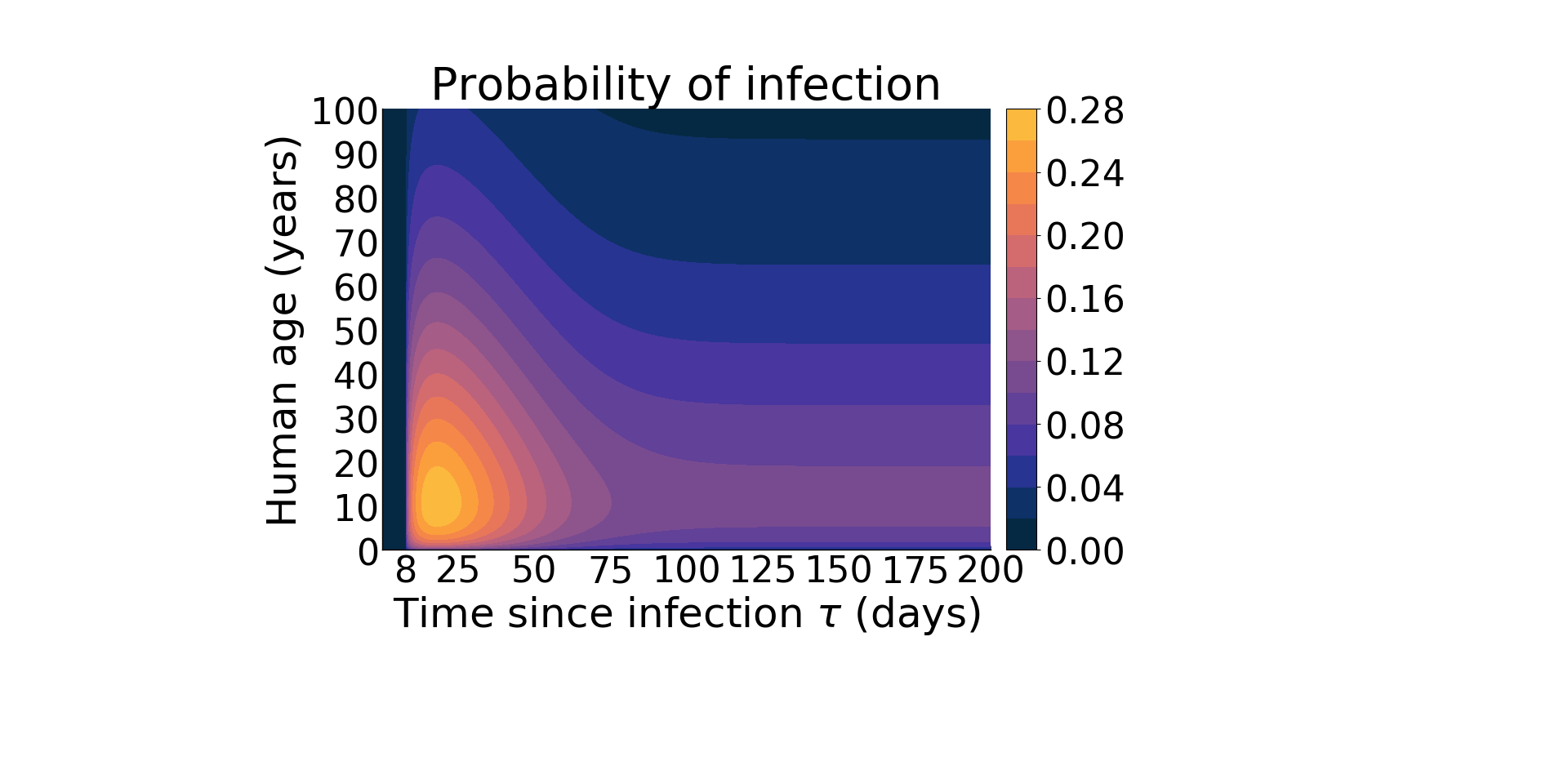}
\caption{The probability that an infected human contaminates a mosquito depending on the time post-infection and the human age.} \label{Fig:ProbaBetaH}
\end{figure}

Note that the function $f$ defined above can be decomposed into the sum of a Gamma distribution with a multiplying factor $x\longmapsto \psi x \exp(-\omega x)$ and an Ivlev function $x\longmapsto \xi(1-\exp(-\omega x))$. While the former function captures the shape of the gametocyte density within the first 40 days after infection given by the data, the latter function captures the limit density as $x$ goes to infinity, since it tends to $\xi$. This is not possible with common distribution law since the function vanishes as x increases. This is particularly important as \textit{Plamodium falciparum} infections may persist for a long time (see \cite{Ashley2014}). We estimate $\xi, \psi$ and $\omega$ respectively to $0.04 [-0.27,0.24]$,  $0.25 [0.23, 0.27]$ and $0.10 [0.086, 0.124]$ for time post-infection larger than 8 days, see Figure \ref{Fig:Gameto_Human} (a). The evolution of the gametocyte density $G$ can then be computed, see Figure \ref{Fig:Gameto_Human} (b). Finally, the probability $\beta_h(a,\tau)$ of infection from humans with of age $a$ and $\tau$ days post-infection is illustrated by Figure \ref{Fig:ProbaBetaH}.

%$$\beta_{h}(a,\tau)=0.071\times \left[\left(22.7 a\exp(-0.0934 a)\right)\times \left(0.04+(0.25 \tau-0.04)\exp(-0.1 \tau)\right)\right]^{0.302}$$

\subsubsection{Age-dependence of the human transmission capability}

\begin{figure}[!h]
\centering
\begin{tabular}{cc}
(a) & (b)\\
\includegraphics[width=.45\linewidth]{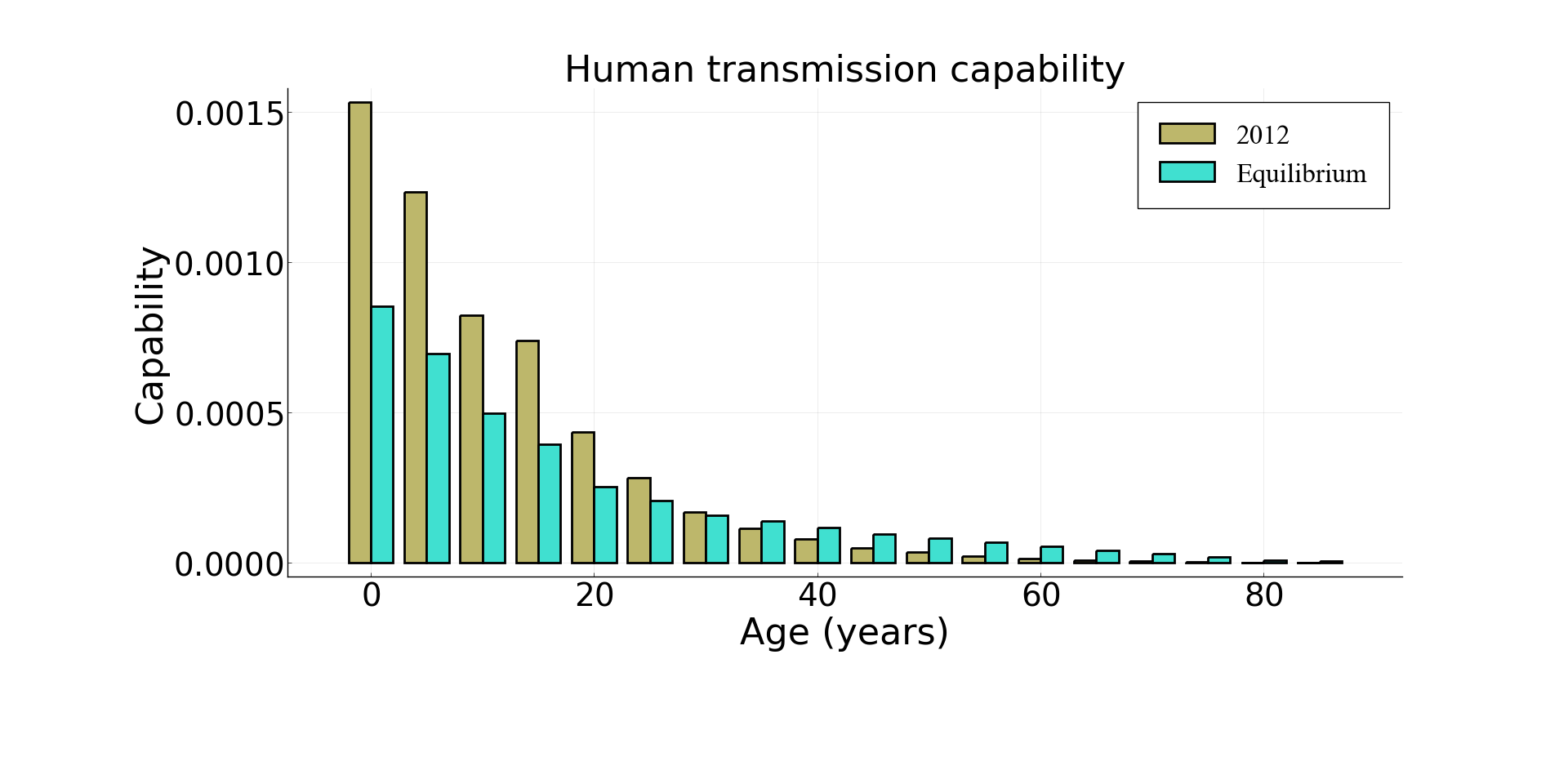} & \includegraphics[width=.5\linewidth]{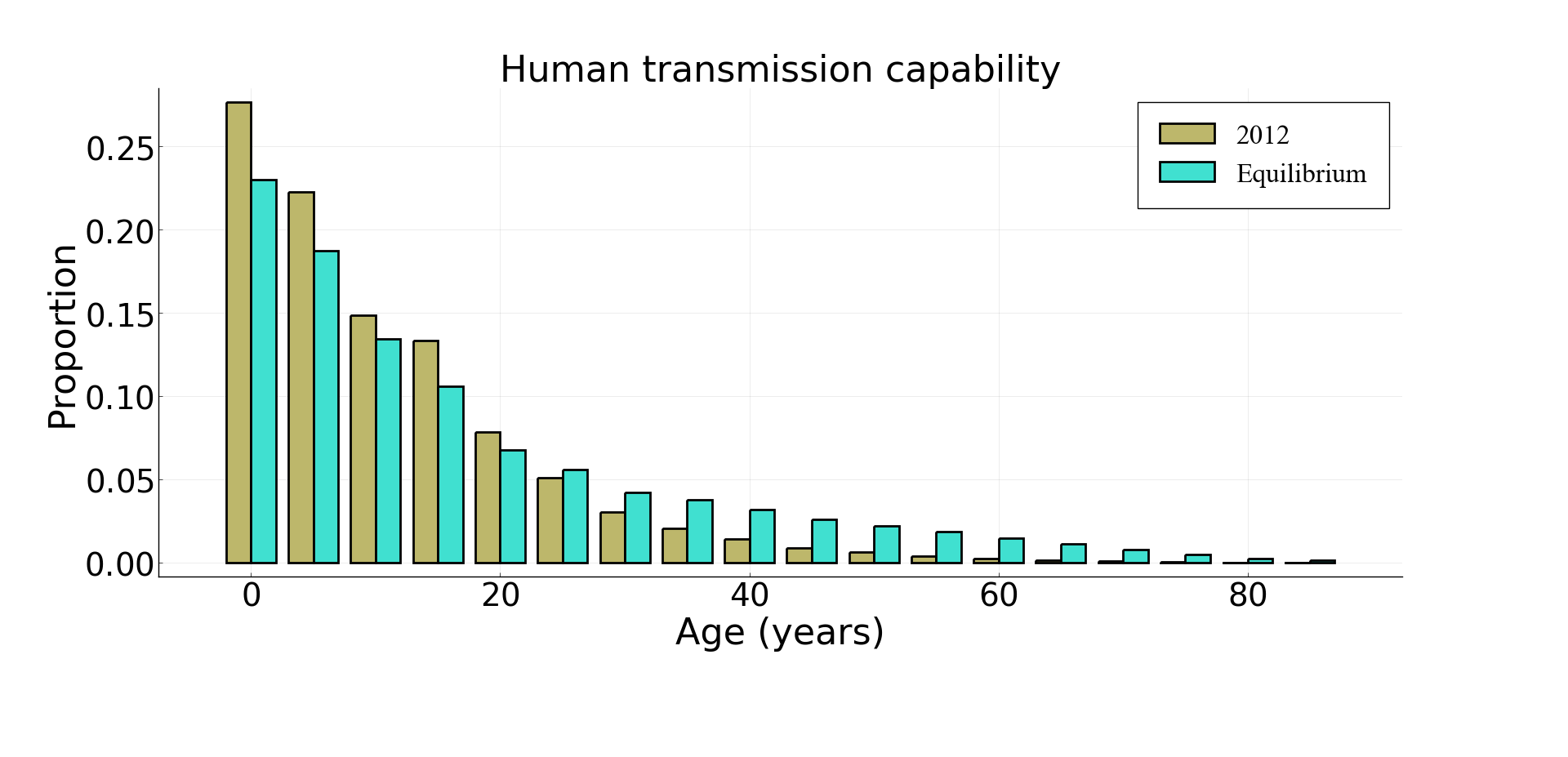}
\end{tabular}
\caption{(a) The contribution of each group of age to the human transmission capability and (b) the proportion for each contribution.} \label{Fig:Human_Capability}
\end{figure}

With these notations and estimates, we see that the human transmission capability $\RR_0^{h\to m}$ can be rewritten according to the probability to remain infected $\pi_h^i$, the human transmission probability $\beta_h$ and the proportion $P_h$ of human of each age as follows:
$$\RR_0^{h\to m}=\int_0^\infty P_h(a)\int_0^\infty \beta_h(a+\tau,\tau)\pi_h^i(a,\tau)d\tau da.$$
In order to see how this quantity depends on the chronological age, we plot the function
$$a\longmapsto P_h(a)\int_0^\infty \beta_h(a+\tau,\tau) \pi_h^i(a,\tau)d\tau$$
(Figure \ref{Fig:Human_Capability} (a)), as well as the proportion for each group of age (Figure \ref{Fig:Human_Capability} (b)). We see that more than half of the human transmission capability comes from the younger population (less than 15 years old). Moreover, we can compare the impact of the change of demographic structure of the human population, mentioned in Section \ref{Sec:HumanDeath}. To this end, we replace $P_h(a)$ by $\frac{S_{h,0}(a)}{\int_0^\infty S_{h,0}(s)ds}$ that was the proportion of humans for each age in 2012 (see Figure \ref{Fig:pop_Bobo} (c)). We see that the capability of people under 30 years old decreases drastically, reducing the total capability by one-third; while the changes in proportion for each group of age are less pronounced.

\subsection{Mosquito transmission capability}

\subsubsection{Mosquito survival}\label{Sec:Mosq_surv}

\begin{figure}[!h]
\centering
\begin{tabular}{cc}
(a) & (b)\\
\includegraphics[width=.45\linewidth]{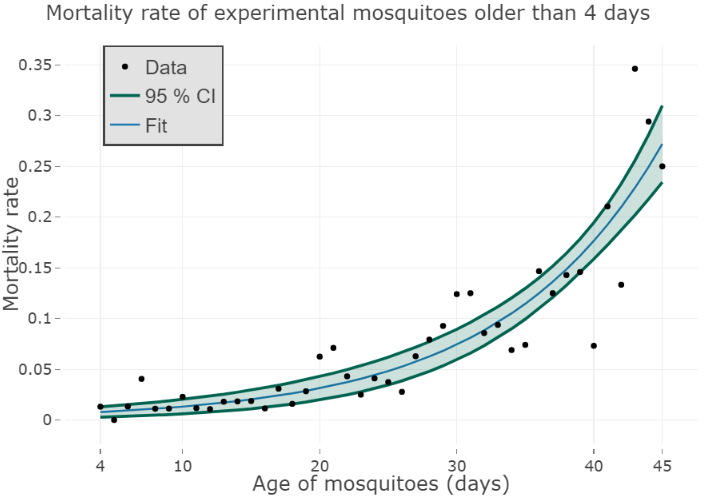} & \includegraphics[width=.45\linewidth]{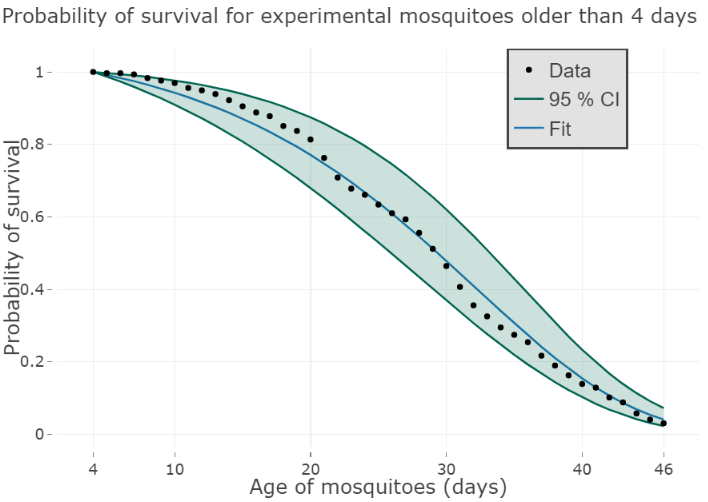}
\end{tabular}
\caption{The mortality rate of experimental mosquitoes older than 4 days is fitted with a Gompertz function by using data from \cite{SomeLefevreGuissou2024} over 295 mosquitoes (a). We then deduce their probability of survival (b).} \label{Fig:morta_mosq_exp}
\end{figure}

As discussed in \cite{Styer2007}, data suggest that mosquitoes do senesce, which means that the assumption of constant mortality rate for mosquitoes, widely used in the malaria modelling literature, is not realistic and may lead for example to an underestimation of the effectiveness of insecticide-treated nets \cite{Iacovidou2022}. While different shapes of functions (Gompertz, logistic) were compared in \cite{Styer2007}, the logistic function were thereafter considered in \cite{Rock2015} in their age-structured model. Here, we make use of the data collected in a recent work \cite{SomeLefevreGuissou2024} over 295 mosquitoes of age between 4 and 46  (we note here that we neglected the death of the 10 last mosquitoes which occurred after day 46 to have better estimates). The mortality rate $\tilde{\mu}_m(a)$ of uninfected mosquitoes of age $a$ for this experiment is such that (Figure \ref{Fig:morta_mosq_exp} (a))
\begin{equation*}
    \tilde{\mu}_m(a)=\begin{cases}
                c_1 & \textnormal{if } a\in[0,4], \\
                c_1 e^{c_2 (a-4)}& \textnormal{if } a\in[4,45],\\
                c_1 e^{41 c_2} & \textnormal{if } a\geq 45,
            \end{cases}
\end{equation*}
with $ c_1\approx 7.85\times 10^{-3} [0.0036,0.0144]$ and $c_2\approx 8.65 \times 10^{-2} [0.0693, 0.1076]$ where we considered constant mortality rate below 4 days old and above 45 days old. In Figure \ref{Fig:morta_mosq_exp} (b) we represented the survival probability after 4 days, that is the function $[4,45]\ni a\longmapsto \exp(-\int_4^a \tilde{\mu}_m(\sigma)d\sigma)$.

\begin{figure}[!h]
\centering
\begin{tabular}{cc}
(a) & (b)\\
\includegraphics[width=0.5\linewidth]{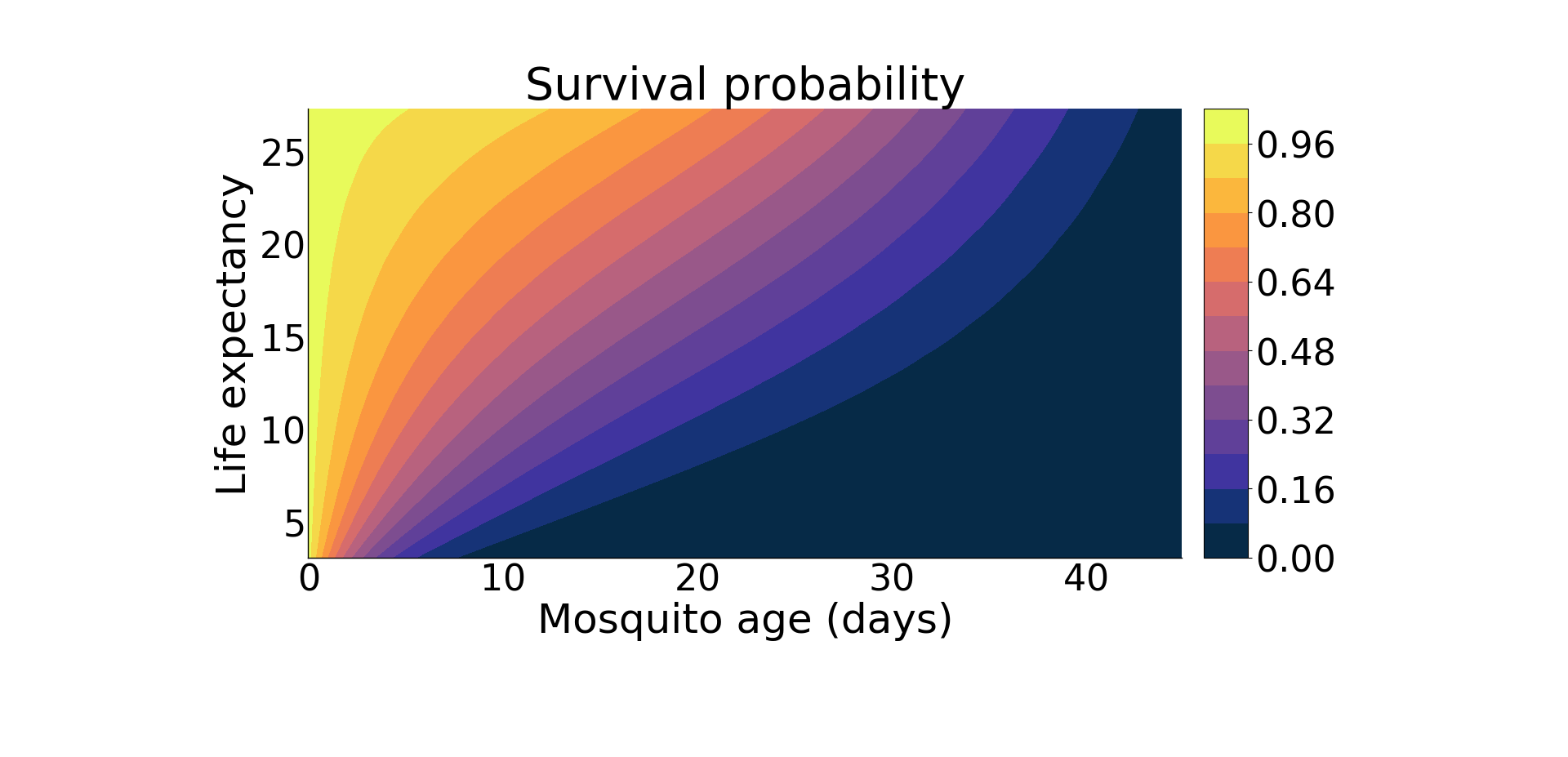} & \includegraphics[width=0.5\linewidth]{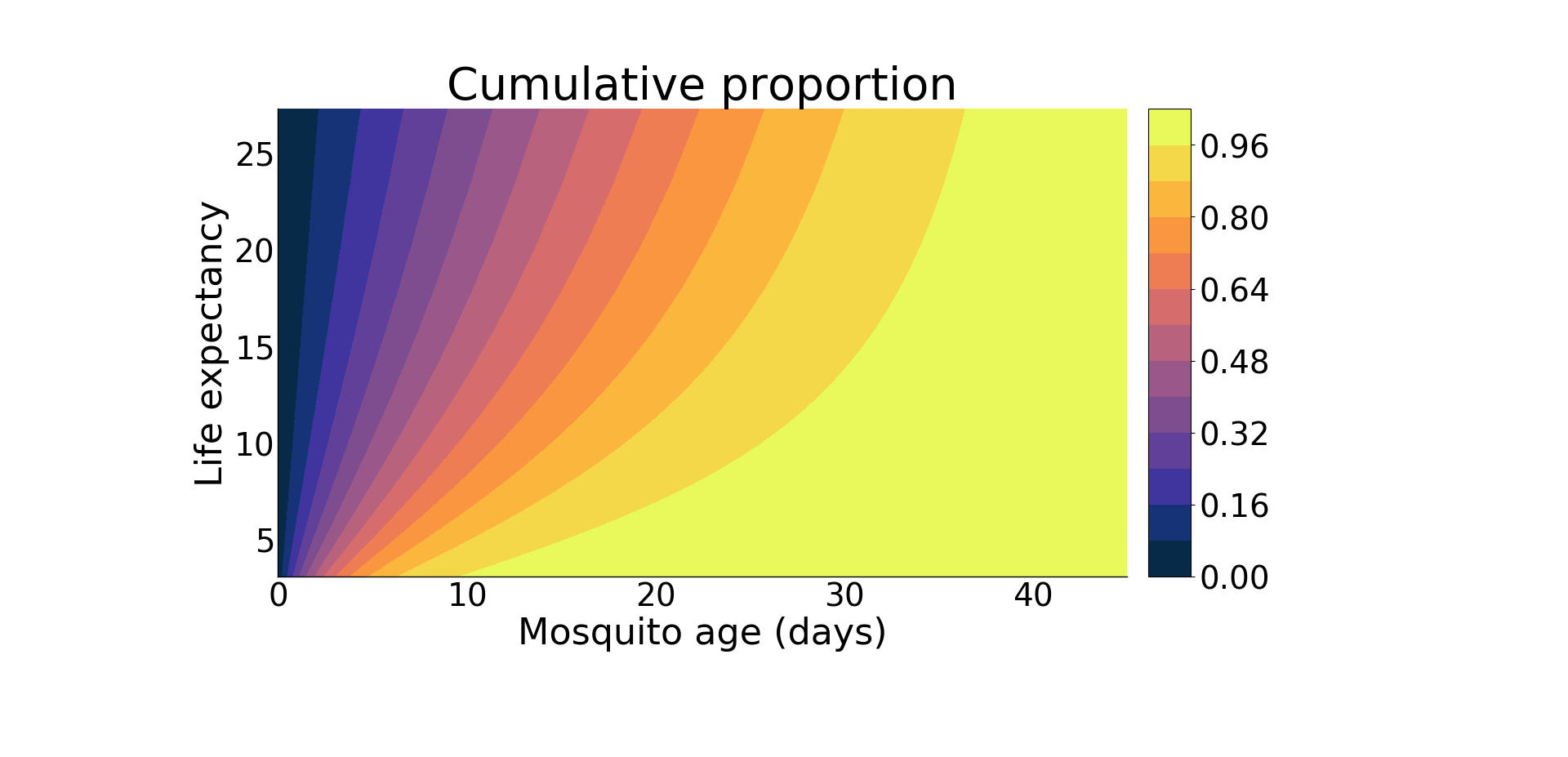}
\end{tabular}
\caption{(a) The survival probability of uninfected mosquitoes according to the chronological age and for different values of $\mu_{\wild }$ resulting on different values of life expectancy. (b) The cumulative sum of the proportion $\int_0^a P_m(s)ds$ of mosquitoes for each chronological age according to the life expectancy (in days).} \label{Fig:morta_mosq_wild}
\end{figure}

With such an experimental mortality rate $\tilde{\mu}_m$, the mean life expectancy is about $27.4$ days corresponding to laboratory conditions. However, the mean life expectancy of wild mosquitoes is in practice hard to estimate. While some modelling papers considered a mean life expectancy of 14 days \cite{Rock2015} or even 30 days \cite{CaiMartcheva2017b}, the data actually show a large variability between different genus \cite{Lambert2022} or even between species \cite{Lambert2022,Matthews2020}. Even within the species complex \textit{Anopheles gambiae s.l.}, whose members are major vectors in Burkina Faso, lead to an important variability: between 3.6 and 15.4 days  \cite{ChitnisCushing2008} or between 4.4 and 10.3 days \cite{Lambert2022}. Reasons for this variability include the effects of seasons and temperature  on mosquito survival \cite{Agyekum2021}, predation, exposure to insecticides, and methods of survival estimation \cite{Matthews2020}. Hence, we will adjust the mortality rate $\tilde{\mu}_m$ by considering $$\mu_m=\mu_{\textnormal{wild}}+\tilde{\mu}_m$$
with a varying constant $\mu_{\textnormal{wild}}$, in order to keep the same shape and to have a mean life expectancy of wild susceptible mosquitoes varying between 3 days to 27.4 days that is the experimental life expectancy. We indeed assume that the survival of wildlife mosquitoes is lower than that of experimental mosquitoes for the reasons listed above (season, temperature, insecticides, predation). We can then observe the survival probability $\pi_m^s(a)= \exp{\left(-\int_0^a \mu_m(\sigma) \d \sigma \right)} $ associated to the different mortality rates in  Figure \ref{Fig:morta_mosq_wild} (a). One component of the mosquito transmission capability is the proportion of mosquitoes of age $a$ at the disease-free equilibrium, that is:
$$P_m(a)=\dfrac{\pi_m^s(a)}{\int_0^\infty \pi_m^s(s)ds}.$$
We can observe in Figure \ref{Fig:morta_mosq_wild} (b) the age distribution of the mosquitoes where almost two-third are less than the life expectancy for each value. 

%resulting to the importance of the vectorial capacity of the young mosquitoes as we shall see later.

\subsubsection{Survival probability of infected mosquitoes}\label{Sec:Mosq_RemInfected}

\begin{figure}[!h]
\centering
\begin{tabular}{cc}
(a) & (b)\\
\includegraphics[width=.4\linewidth]{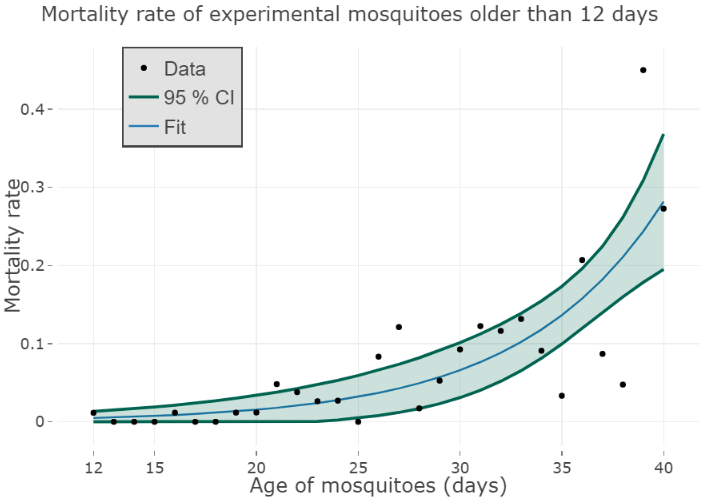} & \includegraphics[width=.4\linewidth]{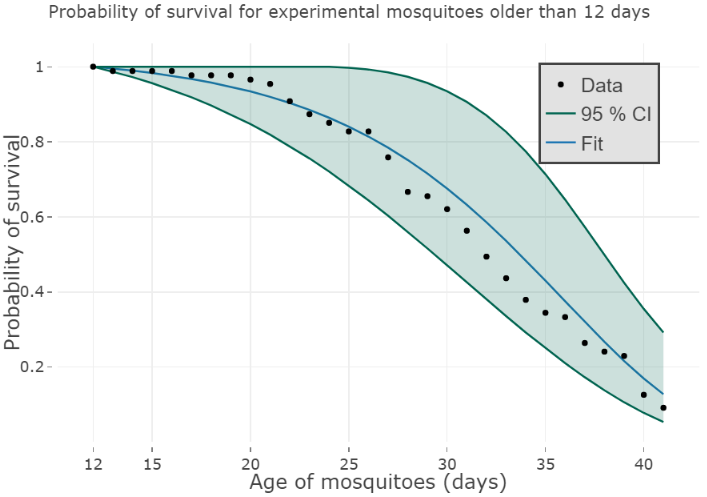}
\end{tabular}
\caption{The mortality rate of experimental infected mosquitoes older than 12 days is fitted with a Gompertz function by using data from \cite{SomeLefevreGuissou2024} over 87 mosquitoes (a). We then deduce their probability of survival (b).} \label{Fig:morta_infec_mosq_exp}
\end{figure}

\begin{figure}[!h]
\centering
\begin{tabular}{cc}
(a) & (b) \\
\includegraphics[width=.45\linewidth]{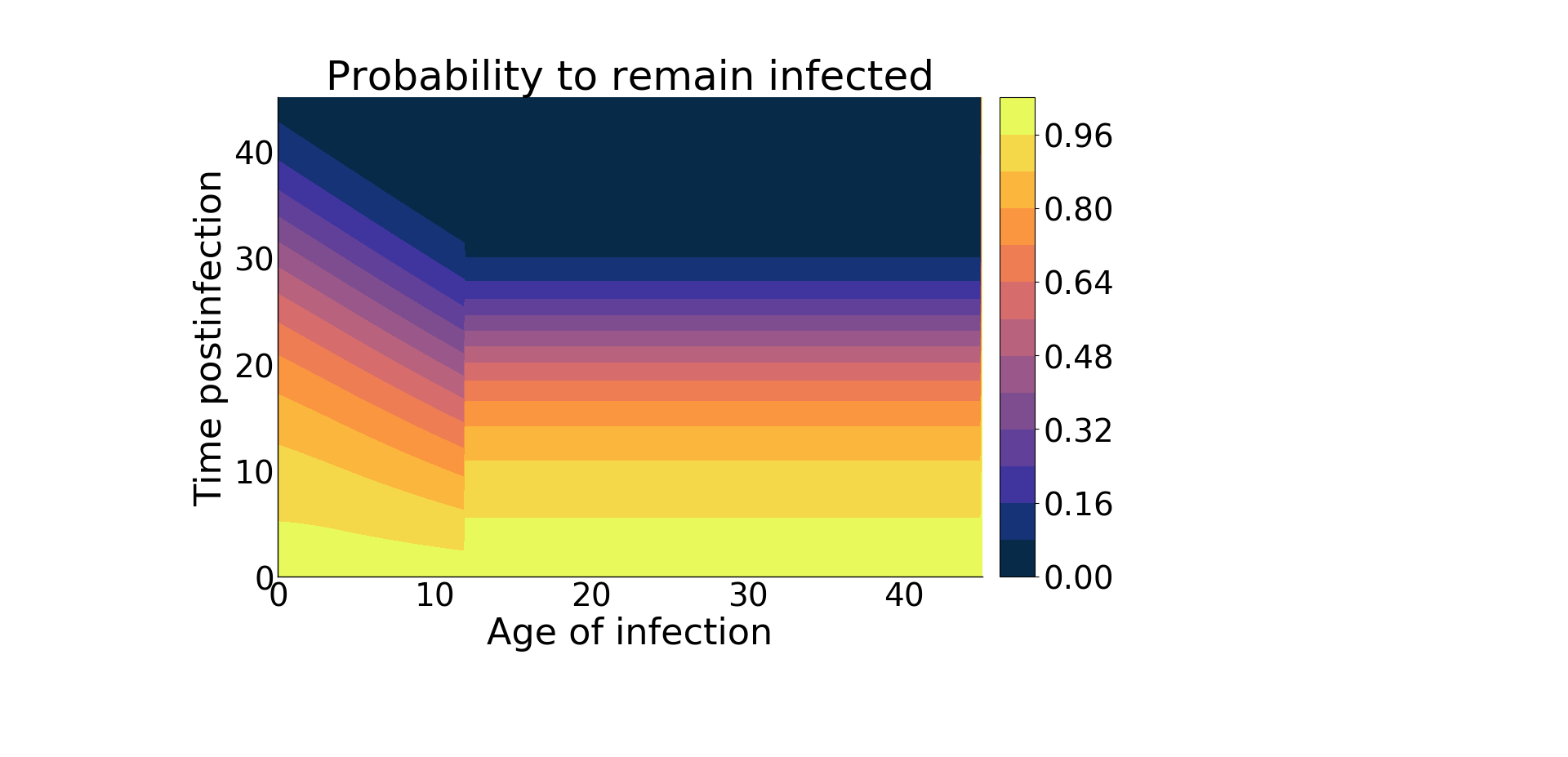} & \includegraphics[width=.45\linewidth]{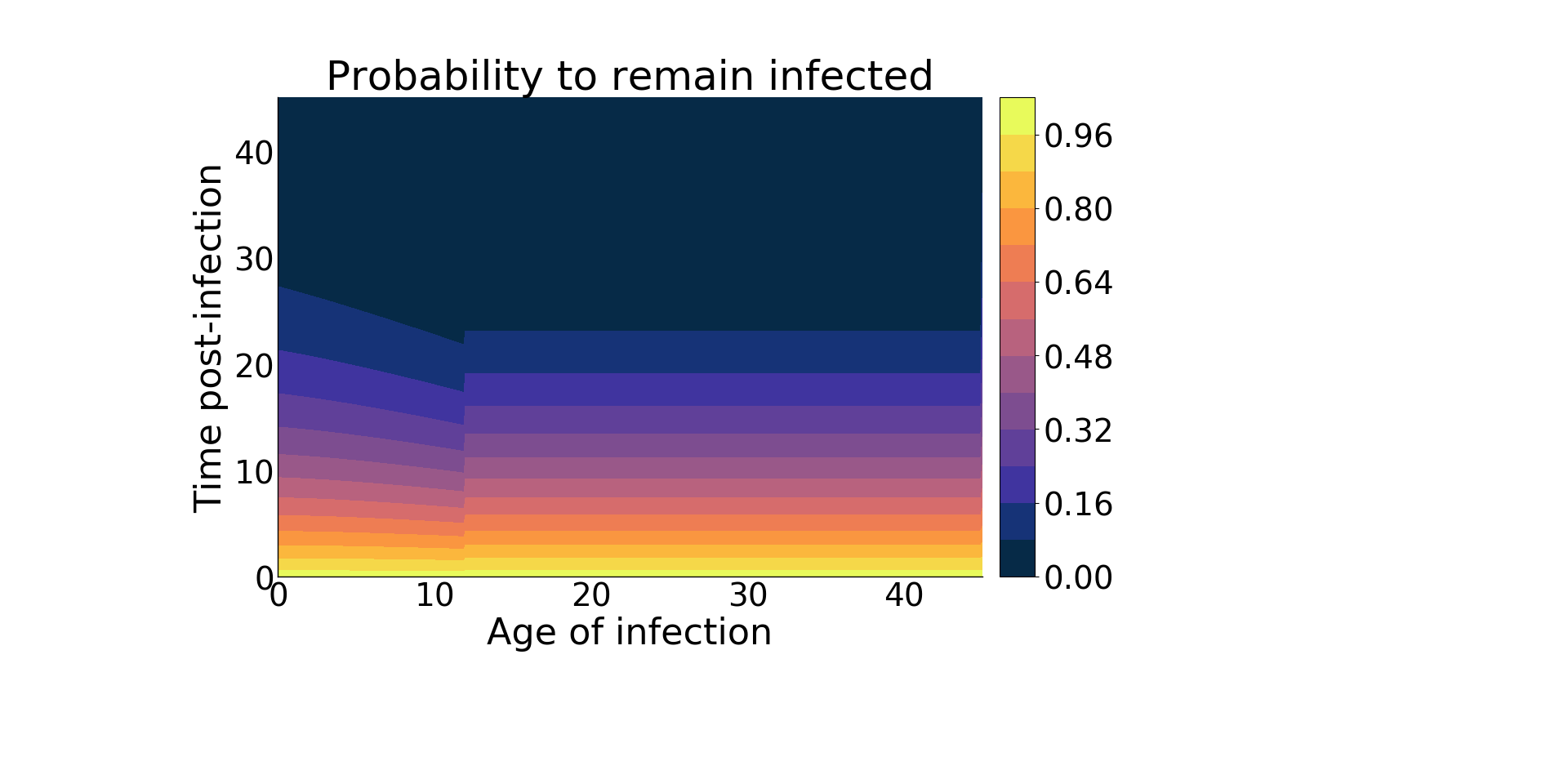}
\end{tabular}
\caption{The probability for the mosquitoes to remain infected $\pi_m^i$ is represented for (a) $\mu_{\wild }=0$ and (b) $\mu_{\wild }=0.07$ for a life expectancy respectively of $27.4$ and $11.4$ days. The age of infection is in days.} \label{Fig:surv_mosq_wild_inf}
\end{figure}

For infected mosquitoes, while the infection does not seem to affect the survival of mosquitoes infected when they are young (4 days old) or middle-aged (8 days old), it seems to have a significant effect on the old mosquitoes survival \cite{SomeLefevreGuissou2024}. Consequently, the experimental mortality rate $\tilde{\nu}_m(a,\tau)$ of an infected mosquito of age $a$ and $\tau$ days post-infection is such that (Figure \ref{Fig:morta_infec_mosq_exp} (a))
\begin{equation*}
\tilde{\nu}_m(a,\tau)=\begin{cases}
\tilde{\mu}_m(a) &\text{if }  a\geq \tau, \quad a-\tau<12, \\
c_1 e^{c_2 \tau} &\text{if }  a\geq \tau, \quad a-\tau\geq 12, \quad \tau\leq 29 \\
c_1 e^{29 c_2} &\text{if }  a\geq \tau, \quad a-\tau\geq 12, \quad \tau> 29 \\
0 &\text{ otherwise}
\end{cases}
\end{equation*}
with $c_1 \approx 4.86 \times 10^{-3} [0.00056, 0.0204]$ and $c_2\approx 1.45 \times 10^{-1} [0.085, 0.253]$. In Figure \ref{Fig:morta_infec_mosq_exp} (b), we represented the survival probability of infected mosquitoes after 12 days, that is the function $[12,45]\ni a\longmapsto \exp(-\int_{12}^a \nu_m(\xi,\xi-12)d\xi)$. As above, we adjust the mortality rate $\tilde{\nu}_m$ by considering
$$\nu_m=\mu_{\textnormal{wild}}+\tilde{\nu}_m.$$
We then define the survival probability of mosquitoes infected at age $a$ for $\tau$ days post-infection by
$$\pi^i_m(a,\tau)=\exp\left(-\int_0^\tau\nu_m(\xi+a,\xi)\d \xi \right)$$
which is equivalent to the probability to remain infected since, once the mosquito salivary glands become invaded by the parasite sporozoites, there is no possible recovery for mosquitoes and they likely remain infectious for life \cite{Guissou2021}. For each time post-infection, we see that the probability decreases as the age at which mosquitoes got infected increases between $0$ and $12$ days, which would not occur with constant mortality rate, though this is less pronounced when the life expectancy decreases. Note here that by life expectancy we refer to the one of susceptible mosquitoes that will never be infected, which is not the one of mosquitoes older than 12 days when they get infected. Indeed, the mosquitoes infected after $12$ days, survive better than the one infected younger. This is due to a lower mortality rate and is discussed in \cite{SomeLefevreGuissou2024}. Nevertheless, the lack of data forced us to suppose that the mortality rate of mosquitoes infected after $12$ days only depends on the time post-infection and not on the chronological age, leading to the same probability of remaining infected, though the probability for the mosquitoes to have survived old enough until being infected may be rather low, depending on the life expectancy (Figure \ref{Fig:surv_mosq_wild_inf}).

 % \comr{Il faut préciser à chaque fois l'unité en âge: days pours time postinfection et days pour age of infection (chez les moustiques) et years quand c'est humain }
\subsubsection{Mosquito to human transmission probability}\label{Sec:Transmission_m}

\begin{figure}[!h]
\centering
\begin{tabular}{cc}
(a) & (b)\\
\includegraphics[width=.44\linewidth]{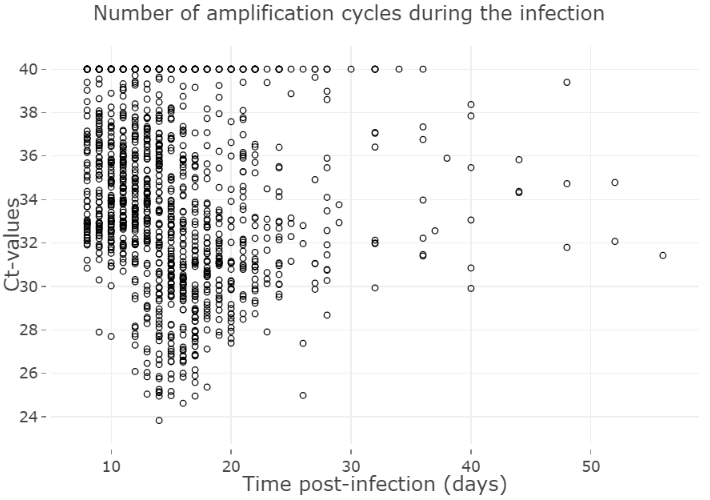} & \includegraphics[width=.44\linewidth]{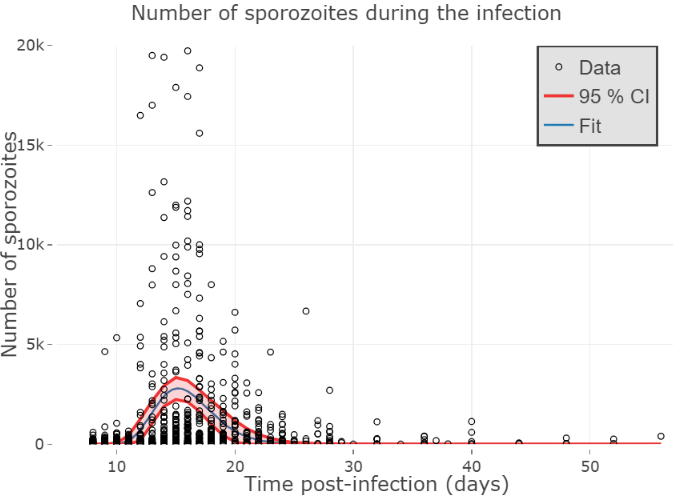}
\end{tabular}   
\caption{(a) Cq-values of infected mosquitoes along the infection and (b) number of sporozoites during the infection.} \label{Fig:Ct-spz}
\end{figure}

\begin{figure}[!h]
\centering
\begin{tabular}{cc}
(a) & (b)\\
\includegraphics[width=.44\linewidth]{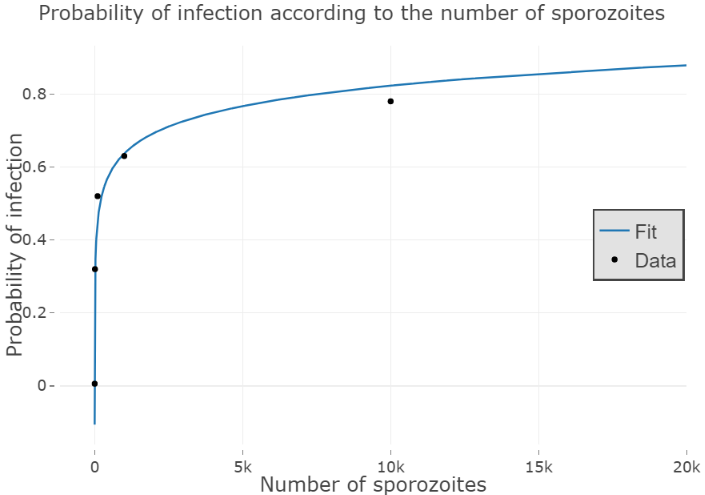} & \includegraphics[width=.44\linewidth]{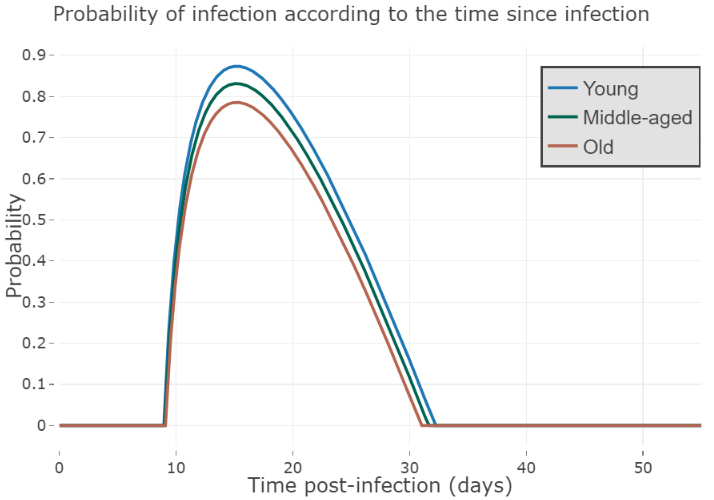}
\end{tabular}
\caption{(a) The probability of infection according to the number of sporozoites in log-scale and (b) the probability of infection $\beta_m$ according to the time post-infection and the age of mosquitoes when infected.} \label{Fig:Proba_Sporo}
\end{figure}

The parasite transmission probability $\beta_m$ from mosquitoes to humans can be computed by using data from \cite{Churcher2017,SomeLefevreGuissou2024}. In \cite{SomeLefevreGuissou2024}, the authors follow a cohort of infected mosquitoes along their infection and get Cq-values, the number of cycles of quantification from qPCR (Figure \ref{Fig:Ct-spz} (a)). The number of sporozoites $\Spz(\tau)$ according to the time since infection $\tau$ is computed by using the following log transformation
\begin{equation}
   \Spz(\tau)=10^{-(C_q(\tau)-40)/3.3} \label{Eq:Spz}.
\end{equation}
This transformation takes into account the maximal number of cycles: 40 (see \cite{SomeLefevreGuissou2024}) leading to a zero probability to have sporozoites for a higher number of amplification cycles. The coefficient 3.3 comes from the negative slope of the linear relationship between the number of cycles of quantitification $C_q$ and the common logarithm of the density of sporozoites \cite{Svec2015} leading to
$$40-3.3 \log_{10}(\Spz(\tau))=C_q(\tau)$$
whence the formula \eqref{Eq:Spz}. We then fitted the density of sporozoites with a gamma distribution (Figure \ref{Fig:Ct-spz} (b)), that is:
\begin{equation}\label{Eq:GammaSpz}
\Spz(\tau)= \begin{cases} 0 &\text{if } \tau\leq 8 \\
C \frac{(\tau-8)^{k-1}\exp\left(-\frac{(\tau-8)}{\xi}\right)}{\Gamma(k)\xi^k} &\text{otherwise } \end{cases}
\end{equation}
where $C$ is the mean number of sporozoites, $\xi$ the rate and $k$ the shape of the gamma distribution. We find with $C\approx 16605 [12873, 21399]$, $k\approx 10.44 [6.18, 17.04]$ and $\xi \approx 0.76 [0.44, 1.42]$. Moreover, in \cite{SomeLefevreGuissou2024}, the authors found that the intensity of sporozoites depends on the age of the mosquito with mean $C_q$-values at 14 days post-infection of 27.46 [27.11,27.81] for 12 day-old mosquitoes, 26.65 [26.33,26.97] for 8 day-old and 25.9 [25.63,26.17] for 4 day-old. We then use the transformation \eqref{Eq:Spz} to rescale the gamma distribution \eqref{Eq:GammaSpz} with the corresponding mean number of sporozoites. Next, according to \cite{Churcher2017}, the link between the number of sporozoites $\Spz$ and the probability of parasite transmission from infected mosquitoes to humans $\beta_m$ is such that $\beta_m = \kappa \log_{10}(\Spz)+\zeta$, where $\kappa\approx 0.186 [0.115, 0.257]$ and  $\zeta\approx 0.08 [0, 0.25]$ (Figure \ref{Fig:Proba_Sporo} (a)). Consequently, we assume that the probability $\beta_m$ is given by the following equation (see Figure \ref{Fig:Proba_Sporo} (b))
\begin{equation}\label{Eq:BetaM}
\beta_m(a,\tau)=
\begin{cases} 0 &\text{if } \tau\leq 8 \text{ or } a-\tau<0 \\
\max(0,0.186 \times \log_{10}\left(\frac{111119(\tau-8)^{9.44}e^{-(\tau-8)/0.76}}{\Gamma(10.44)\times 0.76^{10.44}}\right)+0.08) &\text{if } \tau>8, \ 0\leq a-\tau<8 \\
\max(0,0.186 \times \log_{10}\left(\frac{65842(\tau-8)^{9.44}e^{-(\tau-8)/0.76}}{\Gamma(10.44)\times 0.76^{10.44}}\right)+0.08) &\text{if } \tau>8, \ 8\leq a-\tau<12 \\
\max(0,0.186 \times \log_{10}\left(\frac{37413(\tau-8)^{9.44}e^{-(\tau-8)/0.76}}{\Gamma(10.44)\times 0.76^{10.44}}\right)+0.08) &\text{if } \tau>8, \ a-\tau\geq 12 \end{cases}\end{equation}
Note that the probability is taken as zero for time post-infection less than $8$ days, due to the latent or incubation period within mosquitoes, called the extrinsic incubation period (EIP).

%\begin{align*}
   % \mathcal{K}_{h\to m}(\xi,\tau)=& \beta_{h}(\tau+\xi,\tau) e^{-\int_0^\tau (\nu_h(\sigma+\xi,\sigma)+\gamma_h(\sigma+\xi,\sigma))\d \sigma} \frac{\pi_h(\xi+\tau)}{\int_0^\infty \pi_h(s)\d s},\\
   % \mathcal{K}_{m \to h}(\xi,\tau)=& \beta_m(\tau+\xi,\tau)e^{-\int_0^\tau \nu_m(\sigma+\xi,\sigma)\d \sigma} \frac{\pi_m(\xi)}{\int_0^\infty \pi_m(s)\d s}.
%\end{align*}

%It may be observed that the $\RR _0$ takes the form $$\RR _0^2=V_c \times   \int_0^\infty \int_0^\infty  \mathcal{K}_{h\to m}(\xi,\tau) \d \xi~\d \tau$$
%as the product of the vectorial capacity and the per bite transmission rate from human to mosquito.

\subsubsection{Human feeding rate and mosquito/human ratio}

The human feeding rate, denoted by $\theta$ is defined as the expected number of bites on humans per mosquito per day. An estimation of this rate is complicated to find in the literature and varies for example between 0.13 \cite{Zahar74} and 0.44 \cite{Molineaux79} for \textit{Anopheles gambiae s.l.}. Other estimations can be found in \cite{ChitnisCushing2006} and the references therein. For this reason, we will keep this rate variable.

Another important quantity appearing in the computation of the basic reproduction number is the ratio between mosquitoes and humans. First, the mosquitoes recruitment rate $\Lambda_m$ will be assumed variable. Moreover, while the human mortality rate $\mu_m$ is well-known in Burkina Faso (see Section \ref{Sec:HumanDeath}) and can be assumed not to vary with time, the human recruitment rate $\Lambda_h$ is supposed constant even if the population in Bobo Dioulasso will increase implying a growth of the recruitment rate. Nevertheless, one may assume that this quantity will at some point settle down and give the same age-structure in the demography as shown in Figure \ref{Fig:pop_Bobo} (c). The variability in the parameter $\Lambda_h$ can then be taken into account through another parameter, denoted by $C_{\text{var}}$ such as
$$C_{\text{var}}=\dfrac{\theta^2 \Lambda_m}{\Lambda_h}.$$

\subsubsection{Age-dependence of the mosquito transmission capability}

\begin{figure}[!h]
\centering
\begin{tabular}{cc}
(a) & (b) \\
\includegraphics[width=.45\linewidth]{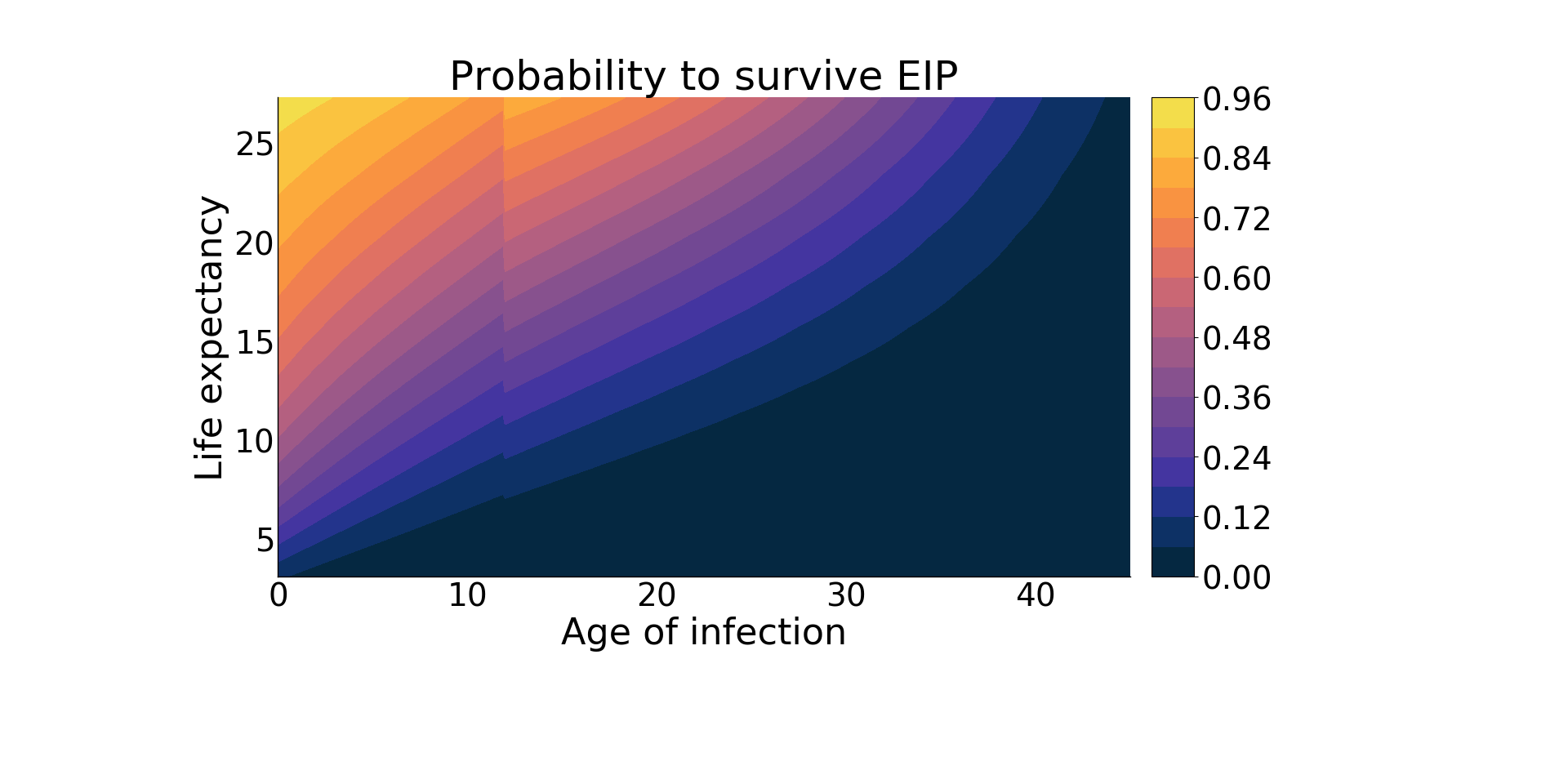}  & \includegraphics[width=.45\linewidth]{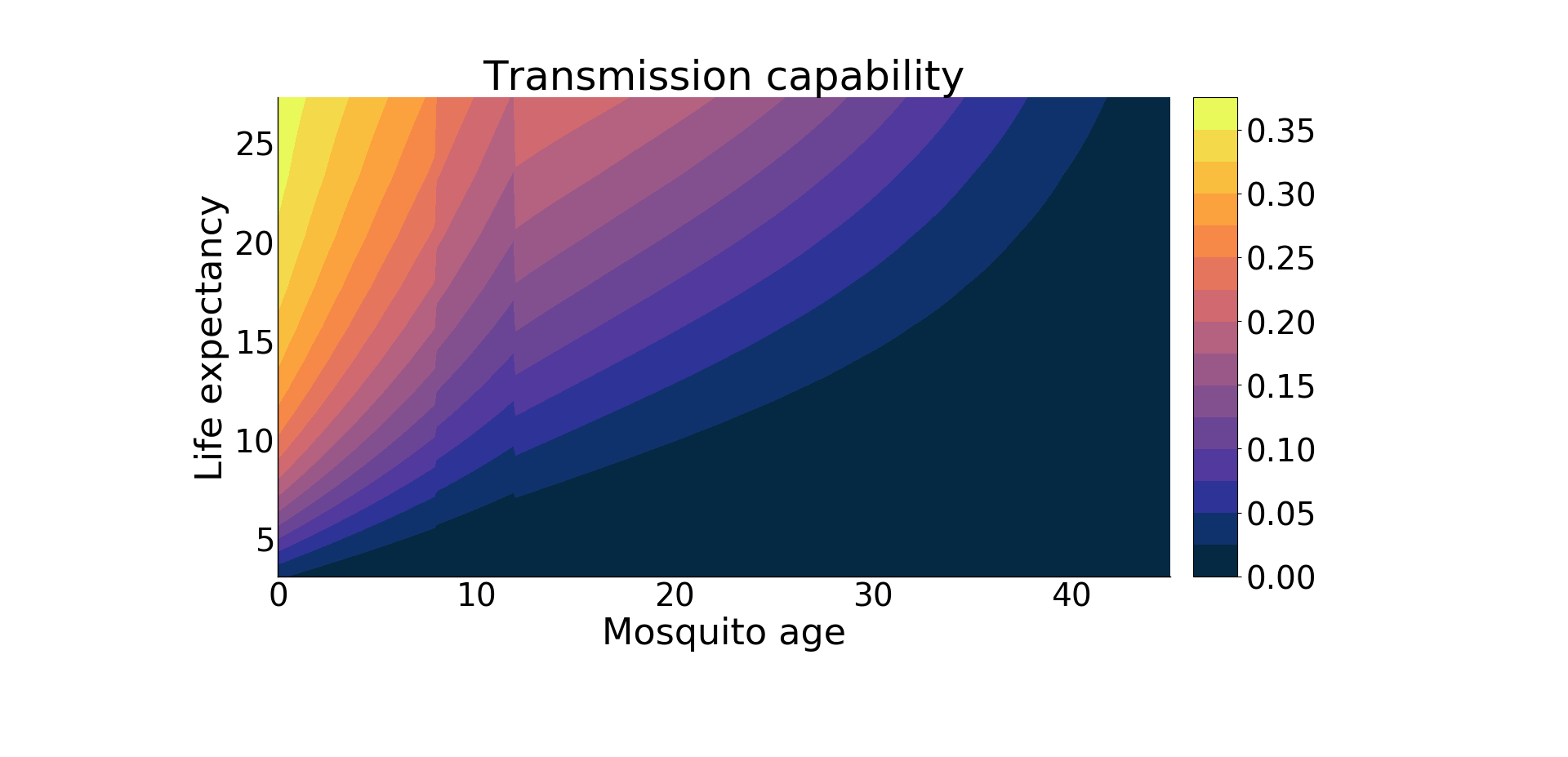}
\end{tabular}
\caption{(a) The probability for mosquitoes to survive the incubation period, according to the life expectancy (in days) and age of infection (in days). (b) The mosquito transmission capability according to the chronological age (in days) and the life expectancy with age-dependant mortality rate.} \label{Fig:Proba_EIP}
\end{figure}

We note here that the variability in the human and mosquito transmissions $\beta_h, \beta_m$ can also be taken into account, up to a multiplying factor, through the constant $C_{\text{var}}$ leading to a mosquito transmission capability $\RR_0^{m\to h}$ that can be rewritten as:
$$\RR_0^{m\to h}=C_{\text{var}}\left(\dfrac{\int_0^\infty \pi_m^s(a)da}{\int_0^\infty \pi_h^s(a)da}\right)\left(\int_0^\infty P_m(a) \int_0^\infty \beta_m(a+\tau,\tau)\pi^i_m(a,\tau)d\tau da \right).$$
Firstly, the mosquito capability transmission requires that the mosquitoes survive the parasite incubation period. For that purpose, we represented on Figure \ref{Fig:Proba_EIP} (a), for each possible age of infection $a$, the probability to survive until age $a$ and also survive the incubation period, \textit{i.e.} $\pi_m^s(a)\times\pi_m^i(a,8)$ according to the life expectancy by varying the parameter $\mu_{\wild}$. As expected, the probability to survive the incubation period increases with the life expectancy (of susceptible mosquitoes) depicting the importance of knowing how long mosquitoes will survive. Also, the older mosquitoes will get infected, the lower is the probability to be someday infectious. Finally, as mentioned in Section \ref{Sec:Mosq_RemInfected}, we observe a slight increase of the probability to survive the incubation period around the age of infection of 12 days. Secondly, decoupling the mosquito transmission capability over the age by drawing the function 
$$a\longmapsto P_m(a)\int_0^\infty \beta_m(a+\tau,\tau)\pi_m^i(a,\tau)d\tau$$
leads to an important part coming from young mosquitoes, since mosquitoes getting infected old will probably never survive the incubation period (see Figure \ref{Fig:Proba_EIP} (b)).

\subsection{On the immunity waning}

The immunity waning regulates the re-entering flux into the susceptible compartment from the recovered one. The basic reproduction number does not depend on the rate $k_h$ since it considers the spread of the disease from a single primary individual into an otherwise susceptible population, whence without recovered individuals. However, as shown in \cite{Richard2021}, the threshold leading to either a backward or a forward bifurcation depends on $k_h$. It results on the importance of parameterizing the latter rate since malaria is endemic in countries such as Burkina Faso. This is hardly understood in practice and different modelling exist in the literature. For example, in \cite{ChitnisCushing2008}, recovered individuals are considered as still infectious and are part of the force of infection. In this case, the rate $k_h$ depicts the loss of the clinical immunity \cite{Keegan2013} that protects against sever symptoms, whence recovered individuals cannot die in their model. As mentioned in \cite{ChitnisCushing2008}, \textit{it is generally believed that immunity is short-lived and requires repeated reinfection to sustain itself}, whence $k_h$ should be age-dependent. In the above papers, it is assumed that the period of immunity is either 3.5 years \cite{Keegan2013} or 5 years \cite{ChitnisCushing2008}. Another example is \cite{Rock2015} where infected humans directly become susceptible instead of being recovered. Last but not least there is a series of papers where the authors developed simple mathematical models by considering the case of acquired immunity boosted by exposure \cite{Aron1983,Aron1988,Aron1988b}.

\section{Derivation of entomological parameters intervening in malaria transmission}\label{Sec:derivation}

The basic reproduction number and the vectorial capacity which were discussed in the latter section, encompass different biological processes (such as mosquito lifespan, human feeding, stability index etc.) that were investigated among others by Smith and McKenzie \cite{SmithEtAl2012,Smith2004,Smith2007}. We derive here formulas for the age-structured model \eqref{Eq:Model} while details are given in Section \ref{Sec:derivation}.

\subsection{Proportion of infected mosquitoes, EIP and sporozoite rate}\label{Sec:Prop_infected}

The force of infection from mosquitoes to humans is defined by \eqref{Eq:lambda_m} and can be rewritten as:
$$\lambda_{m\to h}(t,a)=S_h(t,a)\frac{N_m(t)}{N_h(t)}\int_0^\infty \int_0^\infty \theta \beta_m(s,\tau)\frac{I_m(t,s,\tau)}{N_m(t)}\d s~\d\tau$$
which reveals the mosquito/human ratio $N_h(t)/N_m(t)$ and the proportion of mosquitoes at time $t$, that have chronological age $s$ and infection age $\tau$, which is $I_m(t,s,\tau)/N_m(t)$. Similarly, the force of infection from humans to mosquitoes is defined by \eqref{Eq:lambda_h} and includes the proportion of humans at time $t$, that have chronological age $s$ and infection age $\tau$: $I_h(t,s,\tau)/N_h(t)$. We now assume that the solution of model \eqref{Eq:Model} converges to an endemic equilibrium, \textit{i.e.}
\begin{equation}\label{Hyp:H}\tag{H}
(S_h(t,\cdot), I_h(t,\cdot,\cdot), R_h(t,\cdot,\cdot), S_m(t,\cdot), I_m(t,\cdot,\cdot) \underset{t\to+\infty}{\longrightarrow}  \left(S_h^*(\cdot),I_h^*(\cdot,\cdot), R_h^*(\cdot,\cdot), S_m^*(\cdot), I_m^*(\cdot,\cdot)\right).
\end{equation}
We then denote by $\II_h(a,\tau)$ the ratio of humans of age $a$ that are infected since $\tau$ days:
$$\II_h(a,\tau)=\frac{I_h^*(a,\tau)}{\int_0^\infty S_h^*(a)\d a+\int_0^\infty \int_0^\infty I_h^*(a,\tau)\d a~\d \tau+\int_0^\infty \int_0^\infty R_h^*(a,\tau)\d a~\d\tau}.$$
Note that $\int_0^\infty \int_0^\infty \II_h(s,\tau)\d\tau \d s$ stands for the total proportion of infected humans. We define the probability for a mosquito to survive and get infected (for the first time) at age $a$ by
\begin{equation*}
g_m(a):=\left(\int_0^\infty \int_0^\infty \beta_{h}(s,\tau) \II_h(s,\tau)ds d\tau\right) \theta e^{-\int_0^a \mu_m(s)ds}e^{-\theta a \int_0^\infty \int_0^\infty \beta_{h}(s,\tau) \II_h(s,\tau)ds d\tau}.
\end{equation*}
Summing over all chronological ages, we get the \textbf{probability for a mosquito to ever become infected} as $\int_0^\infty g_m(a)da$, \textit{i.e.}
\begin{equation}\label{Eq:ProbInfected}
1-\int_0^\infty \mu_m(a)e^{-\int_0^a \mu_m(s)ds-\theta a \int_0^\infty \int_0^\infty \beta_{h}(s,\tau) \II_h(s,\tau)ds d\tau}da.
\end{equation}
Concerning the infectious mosquitoes, we first need to define the \textbf{entomological (or extrinsic) incubation period} which is given by
\begin{equation}\label{Eq:EIP}
\EIP :=\inf\left\{\tau \geq 0: \int_0^\infty \beta_m(a,\tau)da>0\right\}
\end{equation}
that corresponds to the smallest infection age $\tau$ so that the probability of malaria transmission from mosquitoes infected since a duration $\tau$ is positive for some chronological ages. In the latter section, it was assumed that $\EIP =8$ as we can observe from \eqref{Eq:BetaM}. For a mosquito infected at age $a$, it follows that the \textbf{probability to survive the incubation period} reads as
\begin{equation}\label{Eq:SurvivingEIP}
e^{-\int_0^{\EIP } \nu_m(a-\tau,\tau)d\tau}
\end{equation}
while the \textbf{probability for a mosquito to ever become infectious} is 
\begin{equation}\label{Eq:ProbInfectious}
\int_{\EIP }^\infty g_m(a-\EIP )e^{-\int_0^{\EIP } \nu_m(a-\tau,\tau)d\tau} da.
\end{equation}
Another important entomological parameter is the \textbf{sporozoite rate} defined as the proportion of mosquitoes that are infectious. This definition assumes that each population is at equilibrium, otherwise the proportion would be time-depending. Computing the \textbf{proportion of mosquitoes that are infected}, we see that this is simply
\begin{equation}\label{Eq:MosqInfected}\frac{C_{\RR } \int_0^\infty \int_{\tau}^\infty \theta e^{-\int_0^{a-\tau} (\mu_m(s)+\theta C_{\RR })ds-\int_0^\tau \nu_m(a-s,s)ds} da d\tau}{\int_0^\infty e^{-\int_0^a (\mu_m(s)+\theta C_{\RR })ds}da+C_{\RR } \int_0^\infty \int_{\tau}^\infty \theta  e^{-\int_0^{a-\tau} (\mu_m(s)+\theta C_{\RR } )ds-\int_0^\tau \nu_m(a-s,s)ds} da d\tau}
\end{equation}
wherein the constant $C_{\RR }$ is given by
$$C_{\RR }:=\left(\int_0^\infty \int_0^\infty \theta \beta_{h}(s,\tau)\II_h(s,\tau)ds d\tau\right).$$
Similarly, the sporozoite rate can be computed as follows:
\begin{equation}\label{Eq:MosqInfectious}\S ^r=\frac{C_{\RR } \int_{\EIP }^\infty \int_{\tau}^\infty \theta e^{-\int_0^{a-\tau} (\mu_m(s)+\theta C_{\RR }  )ds-\int_0^\tau \nu_m(a-s,s)ds} da d\tau}{\int_0^\infty e^{-\int_0^a (\mu_m(s)+\theta C_{\RR })ds}da+C_{\RR } \int_0^\infty \int_{\tau}^\infty \theta e^{-\int_0^{a-\tau} (\mu_m(s)+\theta C_{\RR } )ds-\int_0^\tau \nu_m(a-s,s)ds} da d\tau}.
\end{equation}
 In \cite{Smith2004}, the authors wrote that the sporozoite rate ({\it i.e.}, the proportion of mosquitoes that are infectious) is equivalent to \textit{the probability that an individual mosquito ever become infectious}. But these are two separate issues. One simple counter-example is when the natural mortality is zero ({\it i.e.} $\mu_m\equiv 0$) and the mortality due to the infection is negligible at the beginning of the infection ({\it i.e.} $\nu_m(.,s)\equiv 0 \ \forall s\in [0,\EIP])$ but nonzero otherwise. Then, the probability for any mosquito to become infectious is one (as long as $C_\RR\neq 0$), while the proportion of infected mosquitoes is obviously below $1$ due to the constant flux of newborn susceptible mosquitoes. 

\subsection{Mosquito survival and lifespan}\label{Sec:MosqSurvival}

We now focus on \textbf{mosquito survival}, defined previously by the function $\pi_m^s$. Computing the survival probability of a random mosquito is quite tricky for the general model formulation since it depends on the time of infection. It then requires to know the proportion of mosquitoes that are infected at each age, which was computed in the previous section. Here, $\pi_m^s$ is derived in absence of infection and thus concerns only mosquitoes that will not be infected. As mentioned before, mosquitoes survival is usually supposed not to be affected by malaria infections, \textit{i.e.} $\mu_m=\nu_m$
while our model distinguishes the two mortality rates, whence mosquitoes that will be infected at some point of their lifetime, shall have different survival functions. Using the probability to be infected at each age, we deduce that the probability for a newborn mosquito to survive to age $a$ is:
$$\Pi_m(a)=\pi_m^s(a)e^{-\theta a \int_0^\infty \int_0^\infty \beta_{h}(s,\tau) \II_h(s,\tau)ds d\tau}+
\int_0^a g_m(a-s)e^{-\int_0^s \nu_m(a-\tau,\tau)d\tau}ds$$
while the average lifespan is given by
\begin{equation}\label{Eq:AverageLifespan}
\int_0^\infty a\left[\mu_m(a)\pi_m^s(a)e^{-\theta a \int_0^\infty \int_0^\infty \beta_{h}(s,\tau) \II_h(s,\tau)ds d\tau}+
\int_0^a \nu_m(a,s)g_m(a-s)e^{-\int_0^s \nu_m(a-\tau,\tau)d\tau} ds\right] da.
\end{equation}
The \textbf{median survival time}, \textit{i.e.} the half-life of the mosquito population is defined as the age $\overline{a}$ that satisfies  
\begin{equation}\label{Eq:MedianSurvival}
\Pi_m(\overline{a})=1/2.
\end{equation}
In absence of infection, the half-life of the mosquito population that will never be infected satisfies the following equality:
$$\pi_m^s(\overline{a})=e^{-\int_0^{\overline{a}}\mu_m(s)ds}=1/2 \Longleftrightarrow \int_0^{\overline{a}}\mu_m(s)ds=\ln(2).$$

\subsection{Human feeding, stability index and human blood index}\label{Sec:HumanFeeding}

We now focus on mosquitoes bites, that are the starting point of any infection. It is then crucial to be able to estimate the number of bites per mosquito per day, or equivalently, the number of times each human is bitten every day. With that in mind, we define the \textbf{human feeding rate} as the expected number of bites on humans per mosquito per day. It is here denoted by $\theta$, but is usually decomposed as the product of the mosquito feeding rate $f$ --where the inverse stands for the interval between blood meals-- and the proportion of bites that are taken on humans $\mathcal{Q}$, hence $\theta=f \Q$. For simplicity we supposed that this rate does not depend on the mosquito age, even if our framework would allow any age dependence.

From Section \ref{Sec:MosqSurvival} we know that a mosquito lives on average $\int_0^\infty \Pi_m(a)\d a$ days and bites $\theta$ humans every day. It follows that a mosquito bites in average $\S ^i$ humans over its lifetime, where $\S ^i$ is called the \textbf{stability index} and is defined by
\begin{equation}\label{Eq:StabilityIndex}
\S ^i=\theta\int_0^\infty \Pi_m(a)\d a.
\end{equation}
Since the life expectancy depends both on the age and the infection status, for a mosquito infected at age $\xi$, the average number of bites that it will give after its infection is
$$\theta \int_{\xi}^\infty e^{-\int_{\xi}^s \nu_m(w,w-\xi)dw}\d s,
$$
which is in general not equal to $\S ^i$. The \textbf{human blood index (HBI)} is defined as the proportion of mosquitoes that will ever fed on a human, which we will denote $\H$ in the following. This is given by
\begin{equation}\label{Eq:Mh}
\H=1-\dfrac{\int_0^\infty e^{-\int_0^a(\mu_m(s)+\theta)ds}da}{\int_0^\infty e^{-\int_0^a (\mu_m(s)+\theta C_{\RR })ds}da+C_{\RR } \int_0^\infty \int_{\tau}^\infty \theta e^{-\int_0^{a-\tau} (\mu_m(s)+\theta C_{\RR } )ds-\int_0^\tau \nu_m(a-s,s)ds} da d\tau}.
\end{equation}
Decomposing the human feeding rate as $\theta=f \Q$, one may similarly define the \textbf{proportion of fed mosquitoes}, or the proportion of mosquitoes that have ever fed (whether on human or not) as
\begin{equation}\label{Eq:Hf}
\H_f=1-\frac{\int_0^\infty e^{-\int_0^a(\mu_m(s)+f)ds}da}{\int_0^\infty e^{-\int_0^a (\mu_m(s)+\theta C_{\RR })ds}da+C_{\RR } \int_0^\infty \int_{\tau}^\infty \theta e^{-\int_0^{a-\tau} (\mu_m(s)+\theta C_{\RR }  )ds-\int_0^\tau \nu_m(a-s,s)ds} da d\tau}.
\end{equation}
It is important to report that in the literature (see \textit{e.g.} \cite{Orsborne2018,Vantaux2018}), the human blood index refers to the \textit{proportion of mosquito blood-meals that are of human origin}, which is here denoted by $\Q$.

\subsection{Human biting rate, EIR and lifetime transmission potential}\label{Sec:EIR}

We focus here on infectious bites. Let us define the \textbf{mosquito density per human} as the ratio between the number of mosquitoes and the number of humans, that reads as
\begin{equation}\label{Eq:Ratio_M/H}
\rho=\frac{\Lambda_m\left(\int_0^\infty e^{-\int_0^a (\mu_m(s)+\theta C_{\RR })ds}da+C_{\RR } \int_0^\infty \int_{\tau}^\infty \theta e^{-\int_0^{a-\tau} (\mu_m(s)+\theta C_{\RR }  )ds-\int_0^\tau \nu_m(a-s,s)ds} da d\tau\right)}{\Lambda_h \int_0^\infty e^{-\int_0^a \mu_h(s)ds}da-\int_0^\infty \int_0^a \left(\int_0^s \nu_h(s,\tau)I^*_h(s,\tau)d\tau\right)e^{-\int_s^a \mu_h(\xi)d\xi}ds da}.
\end{equation}
By definition, we know that each mosquito bites $\theta$ times per day. Since the bites does not depend either on the age or its status infection, it follows that the number of bites per human per day, also called the \textbf{human biting rate}, is simply
\begin{equation}\label{Eq:HBR}
\text{HBR}=\rho \theta.
\end{equation}
Since the proportion of infectious mosquitoes is $\S ^r$, it follows that the number of infectious bites received per day by one human, also called the \textbf{entomological inoculation rate (EIR)}, is given by
\begin{equation}\label{Eq:EIR}
    \text{EIR}=\rho \theta \S ^r.
\end{equation}
It is important to note that if each human indeed receives in average $\rho \theta \S^r$ infectious bites per day, each infectious bite has not the same infective load: $\S^r$ is just an indicator for the proportion of infectious mosquitoes without taking into account the time post-infection nor the age, on which depends the transmission rate $\beta_{m}$ (see Section \ref{Sec:Transmission_m}). Furthermore, we may deduce that each mosquito bites in average $\theta \S^r$ humans every day. 

The \textbf{lifetime transmission potential} is then the expected number of new human infections that would be generated by a newly emerged adult mosquito. This is expressed as:
\begin{equation}\label{Eq:transmission_potential}
\left(\theta\int_0^\infty g_m(a)\int_{a}^\infty \beta_m(s,s-a) e^{-\int_a^s \nu_m(\xi,\xi-a)d\xi}ds da)\right)\left(1-\int_0^\infty \int_0^\infty \II_h(a,\tau)d\tau da\right)
\end{equation}
where the first part describes the number of infectious bites given by one mosquito over its lifetime, 
containing the probability to be infected at age $a$ $(g_m(a))$ and the number of infectious bites after infection: $\int_a^\infty \beta_m(s,s-a)e^{-\int_s^a \nu_m(\xi,\xi-a)d\xi}ds$ where the exponential stands for the survival of the mosquito when infected; while the second part is the proportion of susceptible humans. 

\subsection{Comparison to the classical model}\label{Sec:Comparison}

In the previous literature (see {\it e.g.} \cite{Smith2004}) those important entomological parameters were derived in absence of any structural variables. Here we show that our formulas generalize the ones given in \cite{Smith2004} in a particular case, a summary will be given in Table \ref{Table:param}. To this end, we first introduce the classical model, {\it i.e.} without age structure variables by making the following assumption:
\begin{assumption}\label{Assumption}
Parameters $\nu_m, \mu_m, \nu_h, \mu_h, k_h, \gamma_h$ and $\beta_m$ are constant, while $\beta_h$ is constant piecewise function: for each $a\geq 0$,
$$\beta_h(a,\tau)=\begin{cases} 0 & \text{if } \ \tau\leq \EIP \\
\beta_h & \text{otherwise}
\end{cases}
$$
where $\EIP$ is a constant standing for the entomological incubation period. Moreover, the mosquito and human mortality does not depend on the status of the infection: $\nu_m=\mu_m, \ \nu_h=0$.
\end{assumption}
Under Assumption \ref{Assumption}, the PDE model \eqref{Eq:Model} can, after some integration, be rewritten as the following delayed differential system of equations:
\begin{equation}\label{Eq:ModelSimple}
\left\{
\begin{array}{rcl}
S_h'(t)&=&\Lambda_h+k_h R_h(t)-\mu_h S_h(t)-\theta\beta_m I_m(t)\frac{S_h(t)}{N_h(t)} \\
I_h'(t)&=&\theta\beta_m I_m(t)\frac{S_h(t)}{N_h(t)}-\left(\mu_h+\gamma_h\right)I_h(t), \vspace{0.1cm} \\
R_h'(t)&=&\gamma_h I_h(t)-(\mu_h+k_h)R_h(t), \vspace{0.1cm} \\
S_m'(t)&=&\Lambda_m-\mu_m S_m(t)-\theta\beta_h S_m(t)\frac{I_h(t)}{N_h(t)}, \\
E_m'(t)&=&\theta\beta_h S_m(t)\frac{I_h(t)}{N_h(t)}-\mu_m E_m(t)-\theta\beta_h S_m(t-\EIP)\frac{I_h(t-\EIP)}{N_h(t-\EIP)}e^{-\mu_m \EIP}, \\
I_m'(t)&=&\theta\beta_h S_m(t-\EIP)\frac{I_h(t-\EIP)}{N_h(t-\EIP)}e^{-\mu_m \EIP}-\mu_m I_m(t), \\
\end{array}
\right.
\end{equation}
where $S_h$, $I_h$, $R_h$ stand for susceptible, infected and recovered human, while $S_m$, $E_m$, $I_m$ respectively denote susceptible, exposed and infected mosquitoes. We can note that the exposed compartment is composed of the infected mosquitoes whose incubation period is not finished. In the following, \eqref{Eq:ModelSimple} will be referred as the \textit{classical model}. The basic reproduction number $\RR _0$ for the classical model takes the form
\begin{equation}\label{Eq:R0_ODE}
\RR _0= \sqrt{ \RR _0^{m\to h} \times \RR _0^{h\to m}}
\end{equation}
where the per bite transmission capability from human to mosquito $\RR _0^{h\to m}$ and the vectorial capacity $\RR _0^{m\to h}$ now write
\begin{equation}\label{Eq:R0_simple}
\RR _0^{h\to m}= \frac{\beta_h}{\mu_h+\gamma_h}, \qquad \RR _0^{m\to h}= \frac{\theta^2 \Lambda_m  \beta_m \mu_h e^{-\mu_m \EIP}}{\Lambda_h \mu_m^2},
\end{equation}
and we recover the classical formula of the vectorial capacity \cite{MacDonald1957,SmithEtAl2012}. By estimate \eqref{Eq:ProbInfected}, the probability for a mosquito to ever become infected for the classical model writes
$$\frac{\theta \beta_h \II_h}{\mu_m+\theta \beta_h \II_h},$$
where $\II_h$ is the proportion of infected humans (under the static hypothesis \eqref{Hyp:H}). Moreover, by \eqref{Eq:SurvivingEIP}, the probability to survive the incubation period for the classical model becomes $e^{-\mu_m \EIP}$. Finally, by \eqref{Eq:ProbInfectious}, the probability for a mosquito to ever become infectious, for the classical model, is given by $$\frac{\theta \beta_h \II_h}{\mu_m+\theta \beta_h \II_h}e^{-\mu_m \EIP }.$$

By estimates \eqref{Eq:MosqInfected} and \eqref{Eq:MosqInfectious} the proportion of mosquitoes that are infected, for the classical model writes $\frac{\theta \beta_h \II_h}{\mu_m+\theta \beta_h \II_h}.$
Similarly, the sporozoite rate for the classical model writes
$\frac{\theta \beta_h \II_he^{-\mu_m \EIP}}{\mu_m+\theta \beta_h \II_h}.$
Note that in this case, the proportion of mosquitoes that are infected and the sporozoite rate respectively coincide with the probability for each mosquito to become infected and infectious. However, both quantities are expected to be very different in general. Next, based on estimate \eqref{Eq:MedianSurvival}, the median survival time for the classical model, \textit{i.e.} the half-life of the mosquito population leads to $\overline{a}=\frac{ln(2)}{\mu_m}$ which does not depend on the infection status since $\mu_m=\nu_m$.

The human feeding rate (denoted by $\theta$) is the expected number of bites on humans per mosquito per day. It is usually decomposed as the product ($\theta=f \Q$) of the mosquito feeding rate $f$ and the proportion of bites that are taken on humans $\mathcal{Q}$. Note that, for the classical model, the stability index, defined by \eqref{Eq:StabilityIndex}, writes $\S ^i=\frac{\theta}{\mu_m}$. We then observe that the life expectancy of a mosquito that has already lived a certain number of days (infected or not) is exactly the same as a recently emerged mosquito (with the classical formulation), that is $\frac{1}{\mu_m}$ days. As a consequence, Smith and McKenzie \cite{Smith2004} reinterpreted the stability index as the expected number of bites given by a mosquito after it has become infectious. In the general case, this reinterpretation is incorrect. Indeed, the life expectancy depends both on the age and the infection status (see Section \ref{Sec:HumanFeeding}). Moreover, with the classical formulation, the human blood index, defined by \eqref{Eq:Mh}, writes $\H=\frac{\theta}{\mu_m+\theta}$. Similarly, by \eqref{Eq:Hf}, the proportion of fed mosquitoes, or the proportion of mosquitoes that have ever fed (on human or not) for the classical model writes $\H_f=\frac{f}{f+\mu_m}$.

Finally, based on estimates \eqref{Eq:Ratio_M/H}, \eqref{Eq:HBR}, \eqref{Eq:EIR} and \eqref{Eq:transmission_potential}: other quantities (for the classical model formulation) such as mosquito density per human ($\rho$), human biting rate ($\text{HBR}$), entomological inoculation rate ($\text{EIR}$), and lifetime transmission potential are respectively given by $\rho= \frac{\Lambda_m \mu_h}{\Lambda_h \mu_m}$, $\text{HBR}=\frac{\theta\Lambda_m \mu_h}{\Lambda_h \mu_m}$, $\text{EIR}= \frac{\theta^2\Lambda_m \mu_h \beta_h \II_he^{-\mu_m \EIP}}{\Lambda_h \mu_m\left(\mu_m+\theta \beta_h \II_h\right)}$ and 
$\frac{\theta^2 \beta_h \beta_m \II_h e^{-\mu_m \EIP}(1-\II_h)}{\mu_m(\mu_m+\theta \beta_h \II_h)}.$ Note that in \cite{Smith2004}, the second term in the last estimate does not appear, thus neglecting the human already infected. Indeed, new infections imply necessarily upon susceptible individuals.

\begin{table}[!htp]
\begin{center}	\begin{small}
{\renewcommand{\arraystretch}{1.8}\begin{tabular}{|c|c|c|}
 \hline 
			\textbf{Parameters} & \textbf{Classical ODE model} & \textbf{PDE model} \\ \hline
				 Basic reproduction number $(\RR _0)$ & $\sqrt{\frac{\theta^2\Lambda_m \beta_m \beta_h  e^{-\mu_m \EIP}}{\Lambda_h \mu_m^2(\mu_h+\gamma_h)}}$ & \eqref{Eq:R0} \\ \hline
		 Vectorial capacity $(V_c)$ & $\frac{\theta^2 \Lambda_m \mu_h \beta_m  e^{-\mu_m \EIP}}{\Lambda_h \mu_m^2}$ & \eqref{Eq:vec_capacity} \\ \hline
		Probability for a mosquito to become infected	& $\frac{\theta \beta_h \II_h}{\mu_m+\theta \beta_h \II_h}$ & \eqref{Eq:ProbInfected} \\  
		\hline
		 Entomological incubation period $(\EIP)$  & $\EIP ^{(*)}$ & \eqref{Eq:EIP}  \\ 
		 \hline
		 Probability of surviving incubation period & $e^{-\mu_m\EIP }$ & \eqref{Eq:SurvivingEIP}\\ 
		 \hline
		 Probability for a mosquito to become infectious & $\frac{\theta \beta_h \II_h}{\mu_m+\theta\beta_h \II_h}e^{-\mu_m \EIP }$ & \eqref{Eq:ProbInfectious} \\
		 	\hline Proportion of infected mosquitoes	& $\frac{\theta \beta_h \II_h}{\mu_m+\theta \beta_h \II_h}$ & \eqref{Eq:MosqInfected} \\  
				\hline 
		 Sporozoite rate $(\S^r)$ & $\frac{\theta \beta_h \II_h}{\mu_m+\theta\beta_h \II_h}e^{-\mu_m \EIP }$ & \eqref{Eq:MosqInfectious} \\ 		 \hline 
		 Average mosquito lifespan & $\frac{1}{\mu_m}$ & \eqref{Eq:AverageLifespan} \\ \hline
		 Median mosquito survival time & $\frac{\ln(2)}{\mu_m}$ & \eqref{Eq:MedianSurvival} \\ \hline
		 Human feeding rate & $f \Q^{(**)}$ & $\theta$ \\ \hline
		 Stability index $(\S^i)$ & $\frac{\theta}{\mu_m}$ & \eqref{Eq:StabilityIndex} \\ \hline
		 Human blood index $(\H)$ & $\frac{\theta}{\theta+\mu_m}$ & \eqref{Eq:Mh} \\ \hline
		 Proportion of fed mosquitoes $(\H_f)$ & $\frac{f}{f+\mu_m}$ & \eqref{Eq:Hf} \\ \hline
		 %Probability for a mosquito to ever bite a human & $\frac{\theta}{\theta+\mu_m}$ & \eqref{Eq:FedMosqH} \\ \hline
		 %Probability for a mosquito to ever fed & $\frac{f}{f+\mu_m}$ & \eqref{Eq:FedMosq} \\ \hline
		 Density of mosquitoes per human $(\rho)$ & $\frac{\Lambda_m \mu_h}{\Lambda_h \mu_m}$ & \eqref{Eq:Ratio_M/H} \\ \hline
		 Human biting rate $(\text{HBR})$ &  $\frac{\theta\Lambda_m \mu_h}{\Lambda_h \mu_m}$  & \eqref{Eq:HBR} \\ \hline 
		 Entomological inoculation rate $(\text{EIR})$ & $\frac{\Lambda_m \mu_h \theta^2 \beta_h \II_h e^{-\mu_m \EIP }}{\Lambda_h \mu_m\left(\mu_m+\theta\beta_h \II_h\right)}$ & \eqref{Eq:EIR} \\ \hline 
		 Lifetime transmission potential & $\frac{\theta^2 \beta_h \beta_m \II_h e^{-\mu_m \EIP}(1-\II_h)}{\mu_m(\mu_m+\theta \beta_h \II_h)}.$ & \eqref{Eq:transmission_potential} \\
		    \hline
		\end{tabular}}\\
		$^{(*)}$ in the classical model, the EIP is given as a parameter\\
		$^{(**)}$ $f$ stands for the the mosquito feeding rate and $\Q$ is the proportion of bites taken on humans, which are here both aggregated into one parameter: $\theta$
	\end{small}
	\end{center}
	\caption{Summary of the formulas.} \label{Table:param}
\end{table}

\pagebreak

\section{Discussion}

Mathematical models have played an important role in understanding the epidemiology of infectious diseases and particularly about malaria \cite{MacDonald1957,Ross1911}. Quantities such as the basic reproduction number, the vectorial capacity and some important entomological parameters can readily be put into equations once a mathematical framework have been set up. In the previous literature (see \textit{e.g.} \cite{SmithEtAl2012,Smith2004,Smith2007}), those quantities were derived in a classical case for a differential system of equations \eqref{Eq:ModelSimple} \textit{i.e.} without any structural variable, thus neglecting possible age effects. As discussed in Section \ref{Sec:Param}, literature about model parameters clearly show that they depend on both chronological and infection ages of human and mosquito populations. Consequently, this aspect holds significant importance in enhancing our comprehension of malaria transmission and facilitating effective malaria control programs from both human and mosquitoes side. 

From human side, the transmission capability or infectiousness among the human population exhibits significant heterogeneity concerning the age structure (Figure \ref{Fig:Human_Capability}). In general, the transmission capability from an infected human to mosquitoes strongly decreases with the age of the infected individual, with individuals under 30 carrying the majority of the transmission burden (Figures \ref{Fig:Human_Capability} and \ref{Fig:ProbaBetaH}). Despite the inadequacy of data for providing precise quantitative results, such an analysis does enable us to affirm the age-dependence of the human infectious reservoir, primarily among the younger population, especially those under 15 years old (see \cite{Coalson2018,Felger2012}). Demography plays a crucial role, particularly in countries like Burkina Faso, where a significant portion of the population is concentrated in the age range of the major human reservoir (Figure \ref{Fig:pop_Bobo}). 

\begin{figure}[!h]
\centering
\begin{tabular}{cc}
(a) & (b) \\
\includegraphics[width=.45\linewidth]{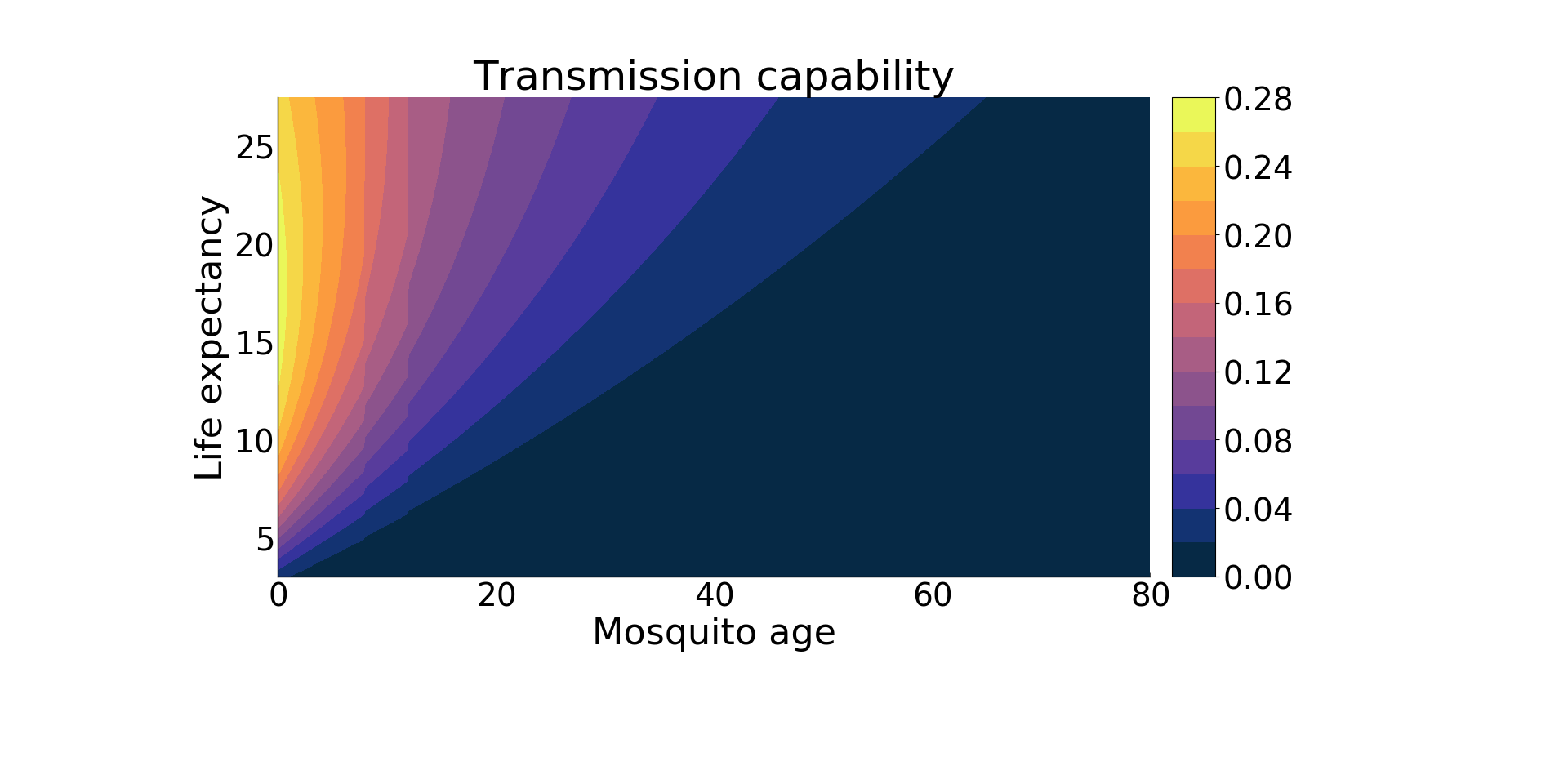} & \includegraphics[width=.45\linewidth]{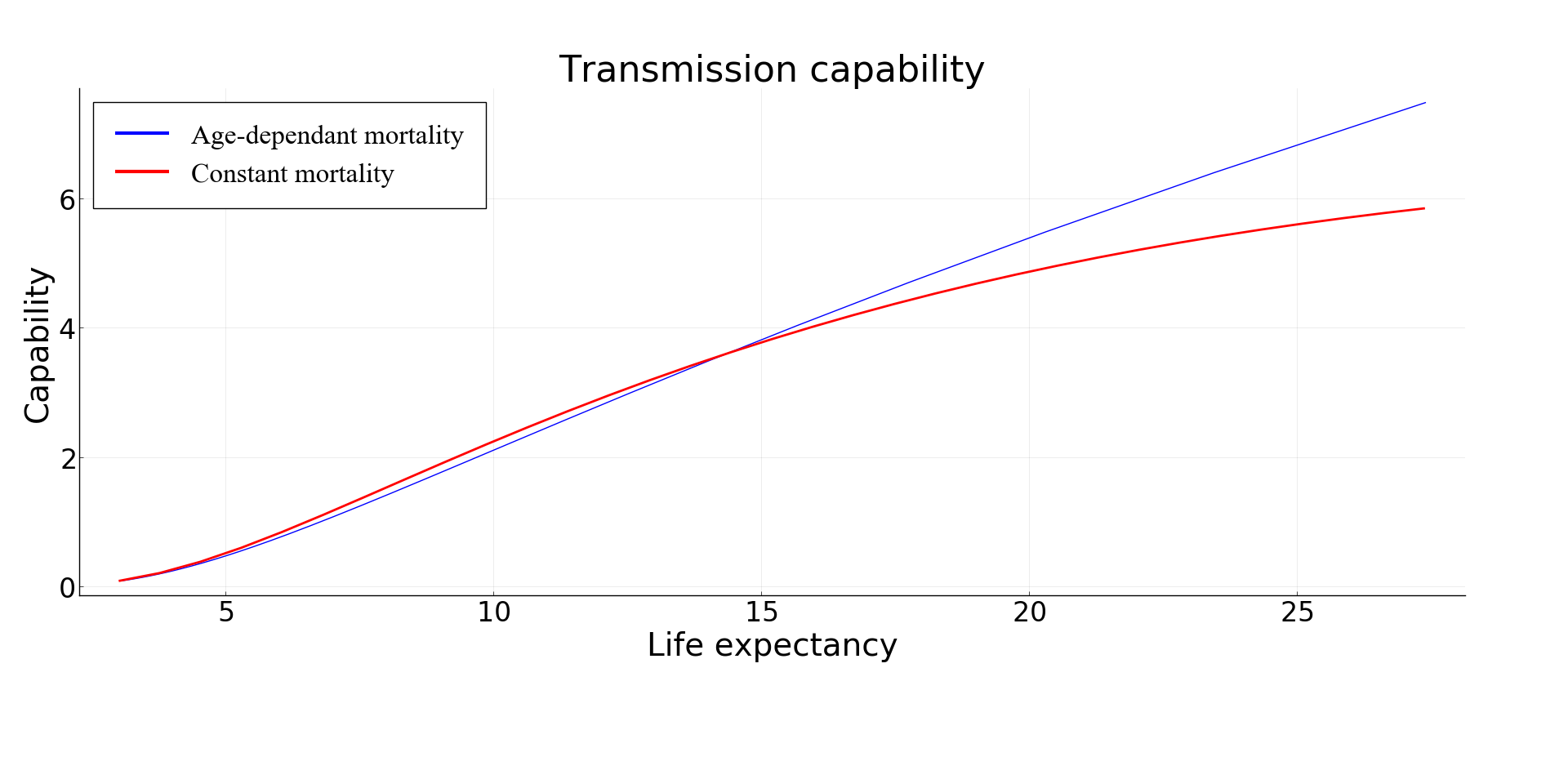}
\end{tabular}
\caption{(a) The mosquito transmission capability according to the chronological age and the life expectancy with constant mortality rate. (b) The mosquito transmission capability summed over the age.} \label{Fig:Mosq_CapabConst}
\end{figure}

\begin{figure}[!h]
\centering
\begin{tabular}{cc}
(a) & (b) \\
\includegraphics[width=.45\linewidth]{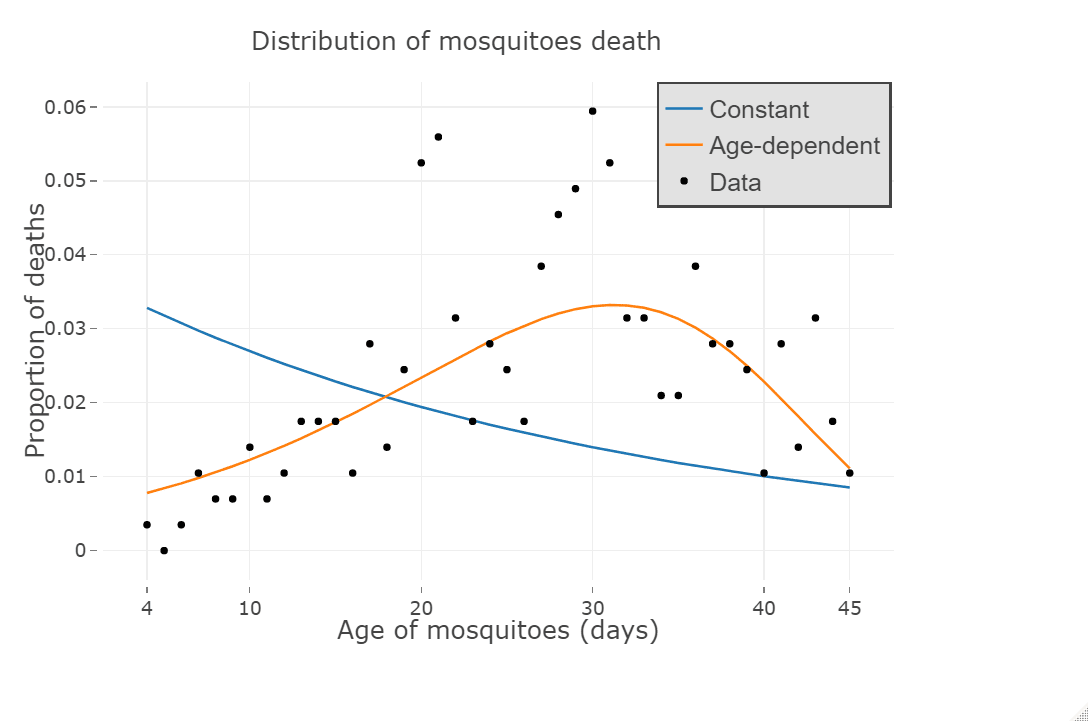} & \includegraphics[width=.5\linewidth]{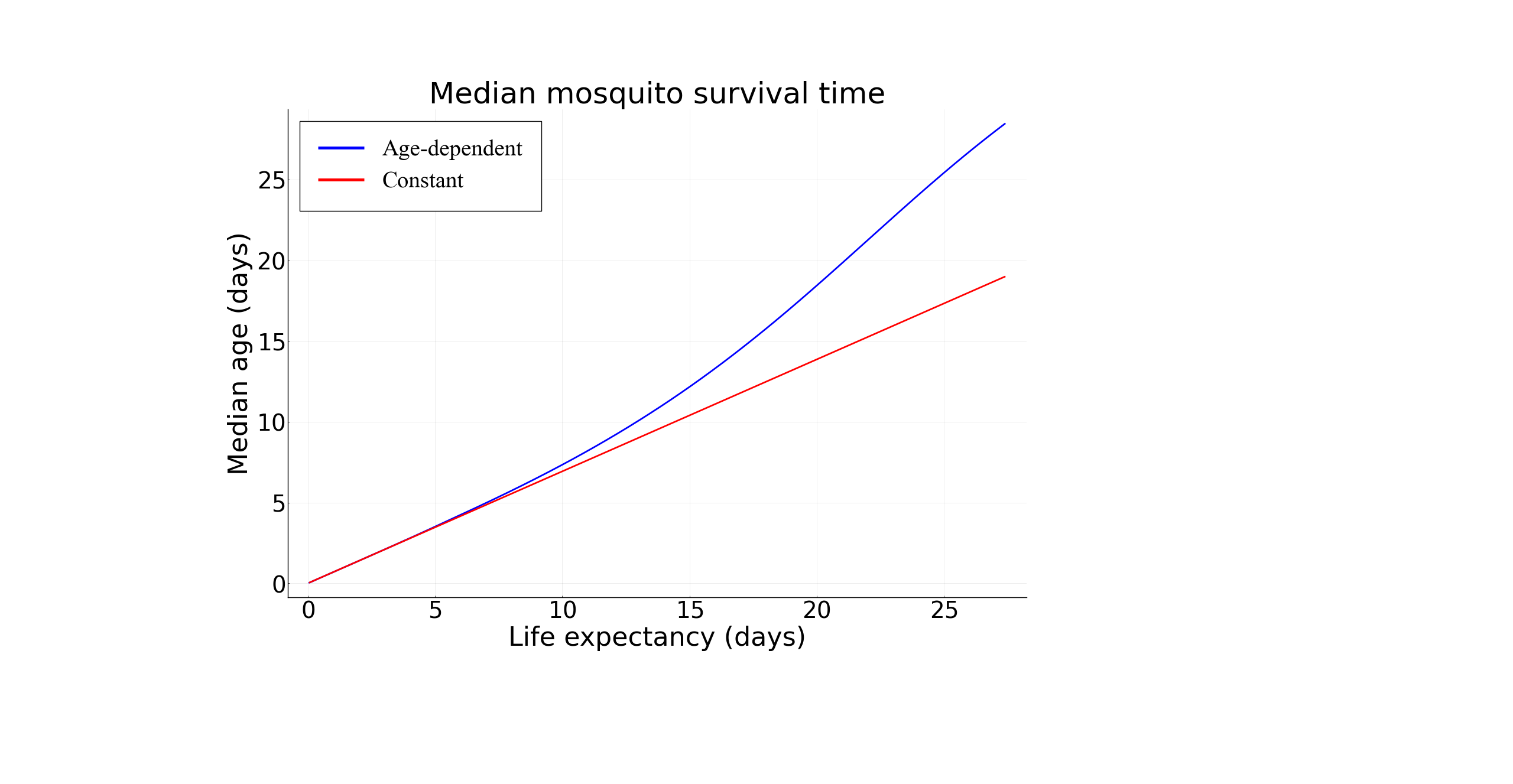} 
\end{tabular}
\caption{Comparison between age-dependent and constant mortality rate for (a) the distributions of mosquitoes death and (b) the median mosquito survival time.} \label{Fig:Mosq_surviv_comp}
\end{figure}

On mosquitoes side, structuring in chronological age seems relevant due to the high dependence of the mortality rate in age depicting a senescence within \textit{Anopheles} populations. While the  survival of mosquitoes in the field contains high variability due to different external (seasons, temperature) and internal (genus, species) reasons as discussed in Section \ref{Sec:Mosq_surv}, the lifespan seems however lower than a few weeks. Adding the fact that mosquitoes do not get infectious right after infection but only after a variable duration called the $\EIP$, we better understand the necessity to follow both chronological ages and infection ages of the mosquito. Indeed, the remaining lifespan of any mosquito depends on the the time already spent, contrary to the classical case. Consequently the sooner a mosquito get infected, the higher its vectorial capacity is (see Figure \ref{Fig:Proba_EIP} (b)). When the mortality is supposed constant, we can observe a similar transmission capability when the life expectancy is lower than 15 days, but differences occur above this value (see Figure \ref{Fig:Mosq_CapabConst} (a)). This is more visible when we compute the total capability that is, up to a multiplying constant, the vectorial capacity:
$$\int_0^\infty P_m(a)\int_0^\infty \beta_m(a+\tau,\tau)\pi_m^i(a,\tau)d\tau da$$
(see Figure \ref{Fig:Mosq_CapabConst} (b)). Here we see that for life expectancy higher than 15 days, the transmission capability is smaller for each mosquito age resulting to a lower total capability. This can be explained by the different age-distribution of mosquitoes at the disease-free equilibrium. In Figure \ref{Fig:Mosq_surviv_comp} (a) we see that the distribution of mosquitoes death, for susceptible mosquitoes, follows well the data with the age-dependent mortality rate while a constant mortality rate overestimates the proportion of mosquitoes that will die young. Consequently, omitting mosquito chronological age in the model may have non negligible epidemiological consequences and implications for malaria control programs. The median mosquito survival time (for susceptibe mosquitoes) also differs with the assumption on the mortality rate, while it is linear with coefficient $\ln(2)$ in the constant case, the age-dependent case shows a nonlinear increase of the median age (see Figure \ref{Fig:Mosq_surviv_comp} (b)). As a consequence, the only knowledge of the mosquito lifespan, through the mortality rate in the classical case, is not enough to quantify epidemiological parameters of malaria transmission.

\section{Appendix}

\subsection{Derivation of formulas of Section \ref{Sec:derivation}}

Here we give some details about the formulas derived in Section \ref{Sec:derivation}. Considering a newborn mosquito, the probability that it will still be alive and not yet infected at age $a$ is
$$\underset{(a)}{\underbrace{e^{-\int_0^a \mu_m(s)ds}}} \  \underset{(b)}{\underbrace{e^{-\theta a \int_0^\infty \int_0^\infty \beta_{h}(s,\tau) \II_h(s,\tau)ds d\tau}}}$$
that is the product of the probability to survive the natural mortality (a) and the probability to not have been infected earlier (b). For such surviving mosquito with age $a$ and not yet infected, the probability that it will get infected at this age is
$$\theta \int_0^\infty \int_0^\infty \beta_{h}(s,\tau)\II_h(s,\tau)ds d\tau.$$
It then follows that the probability for a mosquito to survive then get infected (for the first time) at age $a$ is $g_m(a):=\left(\int_0^\infty \int_0^\infty \beta_{h}(s,\tau) \II_h(s,\tau)ds d\tau\right) \theta e^{-\int_0^a \mu_m(s)ds}e^{-\theta a \int_0^\infty \int_0^\infty \beta_{h}(s,\tau) \II_h(s,\tau)ds d\tau}.$
Summing over all chronological ages, the probability for a mosquito to ever become infected is $\int_0^\infty g_m(a)da$, \textit{i.e.}
$$
\theta\int_0^\infty \int_0^\infty \beta_{h}(s,\tau) \II_h(s,\tau)ds d\tau\int_0^\infty e^{-\int_0^a \mu_m(s)ds-\theta a \int_0^\infty \int_0^\infty \beta_{h}(s,\tau) \II_h(s,\tau)ds d\tau}da
$$
which is equivalent to \eqref{Eq:ProbInfected}. One may observe that $I_m(s,\tau)/N_m$, defined above as the proportion of infected mosquitoes with chronological age $s$ and infection age $\tau$, can readily be computed as:
$$g_m(s-\tau)e^{-\int_0^\tau \nu_m(s-u,u)du}$$
for each $s\geq \tau$ and is zero otherwise. 

The proportion of mosquitoes that get infectious at age $a$ is simply the ones that survived then get infected at age $a-\EIP $ and also survived the incubation period. This writes as
$$g_m(a-\EIP )e^{-\int_0^{\EIP } \nu_m(a-\tau,\tau)d\tau}.$$
It then follows that the probability for a mosquito to ever become infectious is defined by \eqref{Eq:ProbInfectious}.

To compute the proportions of mosquitoes that are infected, we see that this is simply
$$\frac{\int_0^\infty \int_{\tau}^\infty I_m^*(a,\tau)da d\tau}{\int_0^\infty S_m^*(a)da+\int_0^\infty \int_{\tau}^\infty I_m^*(a,\tau)da d\tau }$$
which corresponds to the number of infected mosquitoes over the total number of mosquitoes. Using the method of the characteristics, we readily see that this is equal to \eqref{Eq:MosqInfected}.

Let $\xi$ be the age at which a mosquito is infected. The probability that the mosquito survives until the infection is $\pi_m^s(\xi)$ while the probability to survive from the infection to the age $a$ is $e^{-\int_\xi^a \nu_m(s,s-\xi)d s}$, with $a\geq \xi$. It follows that the probability for a mosquito to survive until age $a$, knowing that it was infected at age $\xi$, is given by
$$\pi_{m,\xi}(a)=e^{-\int_0^\xi \mu_m(s)d s }e^{-\int_\xi^a \nu_m(s,s-\xi)d s }=\pi_m^s(\xi)e^{-\int_\xi^a \nu_m(s,s-\xi)d s }=\pi_m^s(\xi)\pi_m^i(\xi,a-\xi)$$
for each age $a\geq \xi$. Note that if $a<\xi$ then the mosquito dies before being infected and the probability to survive until age $a$ is simply $\pi_m(a)$. We can see that those two terms simply rewrite respectively as $e^{-\mu_m \xi}$ and $e^{-\nu_m (a-\xi)}$ in the classical case. It follows that for \eqref{Eq:ModelSimple} we have $\pi_{m,\xi}(a)=\pi_m^s(a)$ for each $\xi\geq 0$ and consequently the mosquito survival does not depend on the age of infection. For newborn mosquito, the probability to survive until age $a$ is $\Pi_m(a)$
where the left part stands for the survival probability in the susceptible population while the right part is the survival probability in the infected population. One may compute more explicitly this probability in the constant case as:
$$\Pi_m(a)=\begin{cases}
e^{-a(\mu_m+\theta \beta_h \II_h)}\left(1+\frac{\theta\beta_h  \II_h}{\nu_m-\mu_m-\theta \beta_h \II_h}\right)-\frac{\theta \beta_h \II_h e^{-\nu_m a}}{\nu_m-\mu_m-\theta \beta_h \II_h} & \text{ if } \nu_m -\mu_m-\theta \beta_h \II_h \neq 0, \\
e^{-(\mu_m+\theta \beta_h \II_h)a}\left(1+a \theta \beta_h \II_h\right) & \text{ if } \nu_m-\mu_m-\theta\beta_h \II_h=0.
\end{cases}$$
Note that if the mortality of mosquitoes does not depend on the infection \textit{i.e.} $\nu_m\equiv \mu_m$ as in the classical case, or if the proportion of infected humans is zero ($\II_h\equiv 0$), then the latter expression becomes exactly $\pi_m(a)$. Following a newborn mosquito that will never be infected, the probability to survive and die at age $a$ is $\mu_m(a)\pi_m^s(a)$ while the \textbf{average lifespan}, \textit{i.e.} the expected waiting time to death or the mean life expectancy, is defined as
$$\int_0^\infty a\mu_m^s(a)\pi_m(a)da=\int_0^\infty \pi_m^s(a)da=\int_0^\infty e^{-\int_0^a \mu_m(s)ds}da$$
which is equal to $1\frac{1}{\mu_m}$ in the classical case. On the other hand, the average lifespan of mosquitoes that will be infected at age $\xi$, is
$$\int_0^{\xi} a\mu_m(a)\pi_m^s(a)da+\int_{\xi}^{\infty} a\nu_m(a,a-\xi)\pi_{m,\xi}(a)da=\int_0^\infty e^{-\int_0^{\min\{\xi,a\}}\mu_m(s)ds}e^{-\int_{\min\{\xi,a\}}^{a}\nu_m(s,s-\xi)ds}da$$
which amounts to
$$\frac{1}{\mu_m}+e^{-\mu_m \xi}\left(\frac{1}{\nu_m}-\frac{1}{\mu_m}\right)$$
in the constant case that is again $\frac{1}{\mu_m}$ whenever $\mu_m=\nu_m$. We can observe that this quantity is larger than $\frac{1}{\mu_m}$ if and only if $\mu_m>\nu_m$ meaning that infected mosquitoes will live in average longer than susceptible mosquitoes if and only if the mortality rate of the infected population is smaller than the one of the susceptible population. Considering the probability to be infected at each age, we deduce that the average lifespan is given by \eqref{Eq:AverageLifespan} and is equal to
$$\begin{cases}\frac{\mu_m}{(\mu_m+\theta \beta_h \II_h)^2}+\frac{\theta \beta_h \II_h}{\nu_m(\mu_m+\theta \beta_h \II_h)^2}(\mu_m+\theta \beta_h \II_h+\nu_m) & \text{ if } \nu_m-\mu_m-\theta \beta_h \II_h\neq 0, \\
\frac{\mu_m}{(\mu_m+\theta \beta_h \II_h)^2}+\frac{2\theta \beta_h \nu_m \II_h}{(\mu_m+\theta \beta_h \II_h)^3} & \text{ if } \nu_m-\mu_m-\theta \beta_h \II_h=0
\end{cases}$$
in the constant case (or $\frac{1}{\mu_m}$ under Assumption \ref{Assumption}). For mosquitoes that are infected at age $\xi$, the median survival time is defined by the age $\overline{a}$ such as
$$e^{-\int_0^{\min\{\xi,\overline{a}\}}\mu_m(s)ds}e^{-\int_{\min\{\xi,\overline{a}\}}^{\overline{a}}\nu_m(s,s-\xi)ds}=1/2 \Longleftrightarrow \int_0^{\min\{\xi,\overline{a}\}}\mu_m(s)ds+\int_{\min\{\xi,\overline{a}\}}^{\overline{a}}\nu_m(s,s-\xi)ds=\ln(2)$$
which amounts to 
$$\overline{a}=\begin{cases}
\frac{\ln(2)}{\mu_m} & \text{ if } \xi>\frac{\ln(2)}{\mu_m}, \vspace{0.1cm} \\
\frac{\ln(2)+\xi(\nu_m-\mu_m)}{\nu_m} & \text{ if } \xi\leq\frac{\ln(2)}{\mu_m}
\end{cases}$$
in the constant case (and still $\frac{\ln(2)}{\mu_m}$ in the classical case). This distinction can be explained as follows: if the mosquitoes are infected relatively old ($\xi>\frac{\ln(2)}{\mu_m})$ then only the mortality rate of susceptible mosquitoes is relevant for the computations since half of the mosquitoes will be already dead before being infected. On the other hand, if the mosquitoes are infected relatively young ($\xi\leq\frac{\ln(2)}{\mu_m})$ then the computations take also into consideration the mortality rate of infected mosquitoes. We can observe that if $\mu_m\geq \nu_m$ then the median life time $\overline{a}$ is larger than $\frac{\ln(2)}{\mu_m}$, meaning that the infected mosquitoes survive better than the non infected ones. \textit{A contrario}, if $\mu_m\leq \nu_m$ then $\overline{a}\leq \ln(2)/\mu_m$ and susceptible mosquitoes survive better than infected mosquitoes. Let us consider a susceptible mosquito aged of $a$ days, then its life expectancy in absence of infection is
$$\mathcal{E}(a):=\int_a^\infty e^{-\int_a^{s} \mu_m(w)dw}ds.$$
We can note that for a recently emerged mosquito ($a=0$) we recover the average lifespan given earlier. For a mosquito that is aged of $a$ days and that get infected at age $\xi$ (in the past if $a\geq \xi$, or in the future if $a<\xi$), its life expectancy is defined by
$$\mathcal{E}_{\xi}(a):=\int_a^\infty e^{-\int_{\min\{a,\xi\}}^{\xi} \mu_m(w)dw}e^{-\int_{\max\{a,\xi\}}^s \nu_m(w,w-\xi)dw}ds.$$
As a consequence, the number of bites given by a mosquito after it has become infectious, clearly depends on the age at which the mosquito got infected.

In \cite{Smith2004}, the authors defined the human blood index as the proportion of mosquitoes that will ever fed on a human and gave the following formula
$$\H=\frac{\int_0^\infty \pi_m^s(a)(1-e^{-\theta a})da}{\int_0^\infty \pi_m(a)da}$$
where they explained that $\pi_m^s(a)(1-e^{-\theta a})$ is the proportion of mosquitoes that have survived to age $a$ and bitten a human. However, since mosquito survival depends on the infection status, it follows that the latter proportion must depend on when the mosquitoes have bitten a human and whether the bite was infectious or not. It then writes as the ratio between the number of mosquitoes that never have bitten any human, with the total number of mosquitoes. This corresponds to:
\begin{equation*}
\H=1-\frac{\Lambda_m \int_0^\infty e^{-\int_0^a(\mu_m(s)+\theta)ds}da}{\int_0^\infty S_m^*(a)da+\int_0^\infty \int_{\tau}^\infty I_m^*(a,\tau)da d\tau}
\end{equation*}
that is equivalent to \eqref{Eq:Mh}. Similarly, the proportion of fed mosquitoes is defined by
\begin{equation*}
\H_f=1-\frac{\Lambda_m \int_0^\infty e^{-\int_0^a(\mu_m(s)+f)ds}da}{\int_0^\infty S_m^*(a)da+\int_0^\infty \int_{\tau}^\infty I_m^*(a,\tau)da d\tau}
\end{equation*}
which is equivalent to \eqref{Eq:Hf}. We can note that the probability for a newborn mosquito to ever bite a human reads as:
\begin{equation*}\label{Eq:FedMosqH}
1-\int_0^\infty \mu_m(a) \pi_m^s(a) e^{-\theta a}da=\int_0^\infty \theta \pi_m^s(a) e^{-\theta a}da.
\end{equation*}
Indeed, the latter expression can readily be explained as the probability $\pi_m^s(a)$  for the mosquito to survive until age $a$ then to die $\mu_m(a)e^{-\int_0^a \mu_m(s)ds}$ at age $a$ without biting any human $e^{-\theta a}$. As a consequence, the left part is simply the opposite of the proportion of mosquitoes that will never bite any human, because they will die before. Concerning the right part, it can also be decomposed as the probability to survive until age $a$ without never have bitten any human $\pi_m(s)(a) e^{-\theta a}$ and the proportion, among them, that bite a mosquito at the age $a$, that is $\theta$. It is then simply summed over all ages. We can note in the classical case, we recover the same formula as for $\H$, but not in general. In the same way, the probability for a newborn mosquito to ever fed is simply:
\begin{equation*}\label{Eq:FedMosq}
1-\int_0^\infty \mu_m(a)\pi_m^s(a) e^{-f a}da=\int_0^\infty f \pi_m^s(a) e^{-f a}da.
\end{equation*}
We clearly recover the latter formula when mosquitoes only bite humans (\textit{i.e.} $\Q=1$). In the classical case, we see that the latter proportion is simply the same as for $\H_f$. Finally, the ratio between the numbers of mosquitoes and humans is
$$\rho=\frac{\int_0^\infty S_m^*(a)da+\int_0^\infty \int_{\tau}^\infty I_m^*(a,\tau)da d\tau}{\int_0^\infty S_h^*(a)da+\int_0^\infty \int_{\tau}^\infty I_h^*(a,\tau)da d\tau+\int_0^\infty \int_{\eta}^\infty R_h^*(a,\eta)da d\eta}
$$
which leads, after some computations following \cite[Section 7]{Richard2021}, to \eqref{Eq:Ratio_M/H}.

\subsection{Derivation of a SAIR model} \label{Sec:SAIRmodel}

We show here how the model \eqref{Eq:Model} can explicitly highlight the asymptotic stage before infection. Let $A_h(t,a)= \int_0^{\ell} I_h(t,a,\tau) \d \tau $ and $i_h(t,a)= \int_{\ell}^\infty I_h(t,a,\tau) \d \tau $, where $\ell$ is the average duration of asymptomatic stage of an infected individuals. Then $A_h(t,a)$ and $i_h(t,a)$ represent the asymptomatic and symptomatic infections (or clinical cases) respectively. Assume that the probability $\beta_h(s,\tau)$ of parasite transmission from an infected human with age $s$ and infected since a time $\tau$ to any mosquitoes for each bite is such that
$\beta_h(s,\tau)= \beta_h^1(s) 1_{\tau \le  \ell}(\tau) + \beta_h^2(s) 1_{\tau > \ell}(\tau).$ Hence, the force of infection from humans to mosquitoes writes:
\begin{align*}
\lambda_h(t,a)= \frac{1}{N_h(t)} \int_0^\infty \int_0^\infty \theta \beta_h(s,\tau)I_h(t,s,\tau)\d s~\d\tau=  \frac{\theta}{N_h(t)} \int_0^\infty \left(\beta_h^1 (s)A_h(t,s)+ \beta_h^2 (s)i_h(t,s) \right) \d s. 
\end{align*}
Moreover, from \eqref{Eq:Model} we see that
\begin{equation*}
\left\{
\begin{array}{rcl}
\left(\frac{\partial}{\partial t}+\frac{\partial}{\partial a}\right)S_h(t,a)&=&\int_0^\infty k_h(\eta) R_h(t,a,\eta)\d \eta-\mu_h(a) S_h(t,a)-S_h(t,a)\lambda_m(t,a), \\
\left(\frac{\partial}{\partial t}+\frac{\partial}{\partial a}\right)A_h(t,a)&=&I_h(t,a,0) -I_h(t,a,\ell) -\int_0^{\ell} \left(\mu_h(a)+\nu_h(a,\tau)+\gamma_h(a,\tau)\right)I_h(t,a,\tau) \d \tau , \vspace{0.1cm} \\
\left(\frac{\partial}{\partial t}+\frac{\partial}{\partial a}\right)i_h(t,a)&=&I_h(t,a,\ell) -\int_{\ell}^\infty \left(\mu_h(a)+\nu_h(a,\tau)+\gamma_h(a,\tau)\right)I_h(t,a,\tau) \d \tau , \vspace{0.1cm} \\
\left(\frac{\partial}{\partial t}+\frac{\partial}{\partial a}+
\frac{\partial}{\partial \eta}\right)R_h(t,a,\eta)&=&-(\mu_h(a)+k_h(\eta))R_h(t,a,\eta), \vspace{0.1cm} \\
\left(\frac{\partial}{\partial t}+\frac{\partial}{\partial a}\right)S_m(t,a)&=&-\mu_m(a)S_m(t,a)-S_m(t,a)\lambda_{h}(t,a), \\
\left(\frac{\partial }{\partial t}+\frac{\partial }{\partial a}+\frac{\partial }{\partial \tau}\right)I_m(t,a,\tau)&=&-\nu_m(a,\tau)I_m(t,a,\tau).
\end{array}
\right.
\end{equation*}
From \eqref{Eq:Bound_cond}, we deduce that
\begin{equation} \label{eq-Model-SAIRS}
\left\{
\begin{array}{rcl}
\left(\frac{\partial}{\partial t}+\frac{\partial}{\partial a}\right)S_h(t,a)&=&\int_0^\infty k_h(\eta) R_h(t,a,\eta)\d \eta-\mu_h(a) S_h(t,a)-S_h(t,a)\lambda_m(t,a), \\
\left(\frac{\partial}{\partial t}+\frac{\partial}{\partial a}\right)A_h(t,a)&=& S_h(t,a)\lambda_m(t,a) -(\delta_h(a)+\mu_h(a)) A_h(t,a)-\int_0^{\ell} \left(\nu_h(a,\tau)+\gamma_h(a,\tau)\right)I_h(t,a,\tau) \d \tau , \vspace{0.1cm} \\
\left(\frac{\partial}{\partial t}+\frac{\partial}{\partial a}\right)i_h(t,a)&=&\delta_h(a) A_h(t,a) -\mu_h(a)i_h(t,a) -\int_{\ell}^\infty \left(\nu_h(a,\tau)+\gamma_h(a,\tau)\right)I_h(t,a,\tau) \d \tau , \vspace{0.1cm} \\
\left(\frac{\partial}{\partial t}+\frac{\partial}{\partial a}+
\frac{\partial}{\partial \eta}\right)R_h(t,a,\eta)&=&-(\mu_h(a)+k_h(\eta))R_h(t,a,\eta), \vspace{0.1cm} \\
\left(\frac{\partial}{\partial t}+\frac{\partial}{\partial a}\right)S_m(t,a)&=&-\mu_m(a)S_m(t,a)-S_m(t,a)\lambda_{h}(t,a), \\
\left(\frac{\partial }{\partial t}+\frac{\partial }{\partial a}+\frac{\partial }{\partial \tau}\right)I_m(t,a,\tau)&=&-\nu_m(a,\tau)I_m(t,a,\tau), \\
\end{array}
\right.
\end{equation}
where $\delta_h(a):= \dfrac{I_h(t,a,\ell)}{\int_0^{\ell} I_h(t,a,\tau) \d \tau}= \dfrac{I_h(t,a,\ell)}{A_h(t,a)}$ is the rate at which asymptomatic infections progress to symptomatic infections. Next, assume that 
$$
\nu_h(a,\tau)= \nu_h(a) 1_{\tau > \ell}(\tau) \quad \text{and} \quad \gamma_h(a,\tau)=\gamma_{h,1}(a) 1_{\tau \le  \ell}(\tau) + \gamma_{h,2}(a) 1_{\tau > \ell}(\tau).
$$
This assumption implies that death resulting from an infection is negligible for asymptomatic infections. Additionally, the recovery rate for asymptomatic infections is denoted as $\gamma_{h,1}$ while the recovery rate for symptomatic infections is denoted as $\gamma_{h,2}$. Therefore, the latter system ecomes
\begin{equation*}
\left\{
\begin{array}{rcl}
\left(\frac{\partial}{\partial t}+\frac{\partial}{\partial a}\right)S_h(t,a)&=&\int_0^\infty k_h(\eta) R_h(t,a,\eta)\d \eta-\mu_h(a) S_h(t,a)-S_h(t,a)\lambda_m(t,a), \\
\left(\frac{\partial}{\partial t}+\frac{\partial}{\partial a}\right)A_h(t,a)&=& S_h(t,a)\lambda_m(t,a) -(\delta_h(a)+\mu_h(a)+\gamma_{h,1}(a))A_h(t,a), \vspace{0.1cm} \\
\left(\frac{\partial}{\partial t}+\frac{\partial}{\partial a}\right)i_h(t,a)&=&\delta_h(a) A_h(t,a) -( \mu_h(a)+ \nu_h(a)+ \gamma_{h,2}(a))i_h(t,a), \vspace{0.1cm} \\
\left(\frac{\partial}{\partial t}+\frac{\partial}{\partial a}+
\frac{\partial}{\partial \eta}\right)R_h(t,a,\eta)&=&-(\mu_h(a)+k_h(\eta))R_h(t,a,\eta), \vspace{0.1cm} \\
\left(\frac{\partial}{\partial t}+\frac{\partial}{\partial a}\right)S_m(t,a)&=&-\mu_m(a)S_m(t,a)-S_m(t,a)\lambda_{h}(t,a), \\
\left(\frac{\partial }{\partial t}+\frac{\partial }{\partial a}+\frac{\partial }{\partial \tau}\right)I_m(t,a,\tau)&=&-\nu_m(a,\tau)I_m(t,a,\tau), \\
\end{array}
\right.
\end{equation*}
with the following boundary conditions:
\begin{equation*}
\left\{
\begin{array}{rclll}
S_h(t,0)&=&\Lambda_h, &S_m(t,0)=\Lambda_m,\\
A_h(t,0)&=&0, &i_h(t,0)=0, \vspace{0.1cm} \\
R_h(t,a,0)&=& \gamma_{h,1}(a)A_h(t,a) + \gamma_{h,2}(a)i_h(t,a), &R_h(t,0,\eta)=0, \vspace{0.1cm} \\
I_m(t,a,0)&=&S_m(t,a)\lambda_{h}(t,a), &I_m(t,0,\tau)=0.
\end{array}
\right.
\end{equation*}\\

\paragraph{Code availability.}
The code (with R and the Julia Programming Language) used to simulate the model can be accessed through the Zenodo platform at \url{https://doi.org/10.5281/zenodo.10589288}.

% \paragraph{Acknowledgment.}
% The authors would like to thank the anonymous referees for the valuable comments which helped to improve the overall quality of the manuscript.

\bibliography{Bibliography.bib}

\end{document}